\documentclass[preprint2]{aastex}

\newcommand{\myemail}{rjm@astro.caltech.edu}
\newcommand{\rmd}{{\rm d}}

\newcommand{\Map}{M_{\rm ap}}

\newcommand{\etal}{et al.}
\newcommand{\eg}{{\it e.g.}}
\newcommand{\cf}{{\it c.f.}}

\usepackage{amsmath,amssymb}  
\long\def\symbolfootnote[#1]#2{\begingroup 
\def\thefootnote{\fnsymbol{footnote}}\footnote[#1]{#2}\endgroup}
\def\multidrizzle{{\tt MultiDrizzle}}

\slugcomment{For the COSMOS \apj\ special edition}

\shorttitle{3D weak lensing in COSMOS}
\shortauthors{R.\ Massey et al.}

\begin{document}

\title{COSMOS:\\3D weak lensing and the growth of structure}

%
%
%
%
%
\author{Richard Massey\altaffilmark{1\dagger},
Jason Rhodes\altaffilmark{2,1},
Alexie Leauthaud\altaffilmark{3},
Peter Capak\altaffilmark{1}, 
Richard Ellis\altaffilmark{1},
Anton Koekemoer\altaffilmark{4},
Alexandre R\'efr\'egier\altaffilmark{5},
Nick Scoville\altaffilmark{1},
James E. Taylor\altaffilmark{1,6},\\
Justin Albert\altaffilmark{1},
Joel Berg\'{e}\altaffilmark{5},
Catherine Heymans\altaffilmark{7},
David Johnston\altaffilmark{2},
Jean-Paul Kneib\altaffilmark{3}, \\
Yannick Mellier\altaffilmark{8,9},
Bahram Mobasher\altaffilmark{4},
Elisabetta Semboloni\altaffilmark{10,8},
Patrick Shopbell\altaffilmark{1},\\
Lidia Tasca\altaffilmark{3} \&
Ludovic Van Waerbeke\altaffilmark{7}\\
~\vspace{-11mm}
\altaffiltext{1}{California Institute of Technology, 1200 East California Boulevard, Pasadena, CA 91125, U.S.A.}
\altaffiltext{2}{Jet Propulsion Laboratory, Pasadena, CA 91109, USA.}
\altaffiltext{3}{Laboratoire d'Astrophysique de Marseille, BP 8, Traverse du Siphon, 13376 Marseille Cedex 12, France.}
\altaffiltext{4}{Space Telescope Science Institute, 3700 San Martin Drive, Baltimore, MD 21218, U.S.A.}
\altaffiltext{5}{Service d'Astrophysique, CEA/Saclay, 91191 Gif-sur-Yvette, France.}
\altaffiltext{6}{Department of Physics and Astronomy, University of Waterloo, 200 University Avenue West, Waterloo, Ontario N2L3G1, Canada.}
\altaffiltext{7}{Department of Physics \& Astronomy, University of British Columbia, 6224 Agricultural Road, Vancouver, British Columbia V6T 1Z1, Canada.}
\altaffiltext{8}{Institut d'Astrophysique de Paris, UMR7095 CNRS, Universit\'e Pierre~\&~Marie Curie - Paris, 98 bis bd.\ Arago, 75014 Paris, France.}
\altaffiltext{9}{Observatoire de Paris - LERMA, 61 avenue de l'Observatoire, 75014 Paris, France.}
\altaffiltext{10}{Argelander-Institut f\"ur Astronomie, Auf dem H\"ugel 71, D-53121 Bonn, Germany.}
\altaffiltext{$\dagger$}{Email address: {\tt\myemail}\\}
}

\begin{abstract}

We present a three dimensional cosmic shear analysis of the {\it Hubble Space
Telescope} COSMOS survey$^\star$, the largest ever optical imaging program performed
in space. We have measured the shapes of galaxies for the tell-tale distortions
caused by weak gravitational lensing, and traced the growth of that signal as a
function of redshift. Using both 2D and 3D analyses, we measure cosmological
parameters $\Omega_{\rm m}$, the density of matter in the universe, and $\sigma_8$,
the normalization of the matter power spectrum.  The introduction of redshift
information tightens the constraints by a factor of three, and also reduces the relative sampling (or
``cosmic'') variance compared to recent surveys that may be larger but are only two
dimensional. From the 3D analysis, we find 
$\sigma_{8}\left(\Omega_{m}/0.3\right)^{0.44}=0.866^{+0.085}_{-0.068}$ at $68\%$
confidence limits, including both statistical and potential systematic sources of
error in the total budget. Indeed, the absolute calibration  of shear measurement
methods is now the dominant source of uncertainty. Assuming instead a baseline
cosmology to fix the geometry of the universe, we have measured the growth
of structure on both linear and non-linear physical scales.  Our results thus
demonstrate a proof of concept for tomographic analysis techniques that have been
proposed for future weak lensing surveys by a dedicated wide-field telescope in
space.

\end{abstract}


\altaffiltext{$^\star$}{Based on observations with the NASA/ESA {\em Hubble
Space Telescope}, obtained at the Space Telescope Science Institute, which is
operated by AURA Inc, under NASA contract NAS 5-26555; also based on data
collected at: the Subaru Telescope, which is operated by the National
Astronomical Observatory of Japan; the European Southern Observatory, Chile;
Kitt Peak National Observatory, Cerro Tololo Inter-American Observatory, and the
National Optical Astronomy Observatory, which are operated by the Association of
Universities for Research in Astronomy, Inc.\ (AURA) under cooperative agreement
with the National Science Foundation;  the National Radio Astronomy
Observatory, which is a facility of the American National Science Foundation
operated under cooperative agreement by Associated Universities, Inc.; and the
Canada-France-Hawaii Telescope operated by the National Research Council of
Canada, the Centre National de la Recherche Scientifique de France and the
University of Hawaii.}

\section{Introduction}

The observed shapes of distant galaxies become slightly distorted as light from
them passes through foreground mass structures. Such ``cosmic shear'' is induced
by the (differential) gravitational deflection of a light bundle, and happens
regardless of the nature and state of the foreground mass. It is therefore a
uniquely powerful probe of the dark matter distribution, directly and simply
linked to theories of structure formation that may be ill-equipped to predict
the distribution of light \citep[for reviews, see][]{bs,witrev,refrev}.
Furthermore, the main difficulties in this technique lie within the optics of a
telescope that has been built on Earth and can be thoroughly tested. It is not
limited by systematic biases from unknown physics like astrophysical bias
\citep{oferbias,hoebias,smithbias,weinbias} or the mass-temperature
relation for X-ray selected galaxy clusters \citep{hutw,pier,vlnew}.

The study of cosmic shear has rapidly progressed since the simultaneous detection
of a coherent signal by four independent groups \citep{bre,kai00b,wit00,vw00}.
Large, dedicated surveys with ground-based telescopes have recently measured the
projected 2D power spectrum of the large-scale mass distribution and drawn
competitive constraints on cosmological parameters
\citep{browncs,bmer,hamanacs,jarviscs,vw05,xwht,cfhtlsw}. The addition of
photometric redshift estimation for large numbers of galaxies has led to the
first measurements of a changing lensing signal as a function of redshift
\citep{baccombo17,witmanclusters,cfhtlsd}.

The shear measurement methods used for these ground-based surveys have been
precisely calibrated on simulated images containing a known shear signal by the
Shear TEsting Program \citep[STEP;][]{step1,step2}. This program has also sped
the development of a next generation of even more accurate shear measurement
methods \citep{im2shape,shapelets2,bj02,shapelets3,MandelbaumHirata,kkshapelets,
reiko,shapelets4}. With several ambitious plans for dedicated telescopes both on
the ground (\eg\ CTIO-DES, Pan-STARRS, VISTA/VST-KIDS, LSST) and in space (\eg\ 
DUNE, SNAP, and other possible JDEM incarnations), the importance of weak
lensing in future cosmological and astrophysical contexts seems assured.

In this paper, we present statistical results from the first space-based survey
comparable to those from dedicated ground-based observations. The ``cosmic
evolution survey'' \citep[COSMOS;][]{apjse_sco} combines the largest contiguous
expanse of deep imaging from space, with extensive, multicolor follow-up from
the ground. High resolution imaging is particularly needed for weak lensing
because the shapes of galaxies that would also be detected from the ground are
much less affected by the telescope's PSF, and a much higher density of new
galaxy shapes are resolved. This allows the signal to be measured on smaller
physical scales for the first time. Parameter constraints from our survey still
carry a fair deal of statistical uncertainty due to cosmic variance in the
finite survey size; but to a far lesser extent than previous space-based surveys
\citep{rrgmeas,ref02,jr_stis,gems_cs}. More importantly, the potential level of
observational systematics is much lower from space than from the ground, where
the presence of the atmosphere fundamentally limits all weak lensing
measurements.

Extensive ground-based follow-up in multiple filters has also provided
photometric redshift estimates for each galaxy. Lensing requires a purely
geometric measurement, so knowledge of the distances in a lens system as well as
the angles through which light has been deflected are essential. We have
extended cosmic shear analysis into the information-rich three dimensional shear
field. Our constraints on cosmological parameters are tightened by observing
independent galaxies at multiple redshifts, and the separate volume in each
redshift slice reduces the cosmic variance. Furthermore, we can directly trace
the growth of large-scale structure on both linear and non-linear physical
scales. Although these results are still limited by the finite size of the
COSMOS survey, they provide a ``proof of concept'' for tomographic techniques
suggested \citep[by \eg][]{taylor3d,bj3d,heavens3d,taylor3d2} for future
missions dedicated to weak lensing. Throughout this paper, we have assumed a
flat universe, with the Hubble parameter $h=0.7$.


This paper is organized as follows. In \S\ref{sec:method}, we describe the data
and analysis techniques. In \S\ref{sec:2dresults}, we present a traditional 2D
``cosmic shear'' analysis of the two-point correlation functions, demonstrating
the level to which systematic effects have been eliminated from the COSMOS data.
In \S\ref{sec:3dresults}, we extend the analysis into three dimensions via
redshift tomography. We show how the signal grows as a function of redshift, and
directly trace the growth of structure over cosmic time, on a range of physical
scales. In \S\ref{sec:constraints}, we use the measured statistics from both the
2D and 3D analyses to derive constraints on cosmological parameters. We
conclude in \S\ref{sec:conclusions}.


\section{Data analysis methods} \label{sec:method}

\subsection{Image acquisition}

The COSMOS field is a contiguous square, covering 1.64 square degrees and
centered at 10:00:28.6, +02:12:21.0 (J2000) \citep{apjse_sco2,apjse_koe}. Between
October 2003 and June 2005, the region was completely tiled by 575 slightly
overlapping pointings of the {\it Advanced Camera for Surveys} (ACS) {\it Wide
Field Camera} (WFC) with the $F814W$ (approximately $I$-band) filter. Four,
slightly dithered, 507~second exposures were taken at each pointing. Compact
objects can be detected on the stacked images in a $0.15\arcsec$
diameter aperture at $5\sigma$ down to $F814W_{AB}=26.6$ \citep{apjse_sco}.

The individual images were reduced using the standard STScI ACS pipeline, and
combined using \multidrizzle\ \citep{drizzle,apjse_koe}. We took care to optimize
various \multidrizzle\ parameters for precise galaxy shape measurement in the
stacked images \citep{apjse_rho}. We use a finer pixel scale of $0.03\arcsec$
for the stacked images. Pixelization acts as a convolution followed by a
resampling and, although current algorithms can successfully correct for
convolution, the formalism to properly treat resampling is still under
development for the next generation of methods.
We use a Gaussian {\sc
DRIZZLE} kernel that is isotropic and, with \texttt{pixfrac}$=0.8$, small enough to avoid
smearing the object unnecessarily while large enough to guarantee that the
convolution dominates the resampling. This process is then properly corrected by
existing shear measurement methods.

\subsection{Shear measurement}

The detection of objects and measurement of their shapes is fully described in
\citet{apjse_lea}. Modelling of the ACS PSF is discussed in \citet{apjse_rho}.
Here we provide only a brief summary of the important results.

Objects were detected in the reduced ACS images using {\tt SourceExtractor}
\citep{sex}. To avoid biasing our result, the detection threshold was set
intentionally low: far beneath the final thresholds that we  adopt.
The catalog was finally separated into stars and galaxies by noting their
positions on the magnitude {\it vs} peak surface brightness plane. Objects near
bright stars or any saturated pixels were masked using an automatic algorithm,
to avoid shape biases due to any background gradient. The images were then all
visually inspected, to mask other defects by hand (including ghosting, reflected
light and asteroid/satellite trails).

The size and the ellipticity of the ACS PSF varies over time, due to the thermal
``breathing'' of the spacecraft. The long period of time during which the COSMOS data was
collected forces us to consider this effect. Although other strategies have been demonstrated
successfully for observations conducted on a shorter time span, it would be inappropriate for
us to assume, like \citet{lombardi05}, that  the PSF is constant or even, like
\citet{gems_cs}, that the focus is piecewise constant. Fortunately, most of the PSF
variations can be ascribed to a single physical parameter: the distance between the primary
and secondary mirrors or ``effective focus''. Variations of order $10\mu$m create ellipticity
variations of up to $5\%$ at the edges of the field, which is overwhelming in terms of a weak
lensing signal. \citet{jee05} built a PSF model for individual exposures by linearly
interpolating between two PSF patterns, observed above and below nominal focus. We have used
the {\sc TinyTim} \citep{krist03} raytracing package to continuously model the PSF as a
function of effective focus and CCD position. By matching the dozen or so stars brighter than
$F814W_{AB}=23$ on each typical COSMOS image \citep{apjse_lea} to {\sc TinyTim} models, we
can robustly estimate the offset from nominal focus with an rms error less than $1\mu$m
\citep{apjse_rho}. We then return to the entire observational dataset, and fit a $3\times
2\times 2$ order polynomial for each parameter of the PSF model, as a function of $x$, $y$,
and focus. Using the entire COSMOS dataset strengthens the fit, especially at the extremes of
used focus values, where few stars have been observed. The final PSF model for each exposure
is then extracted from the 3D fit, at the appropriate focus value.

We use the shear measurement method developed for space-based imaging by
\citet[][hereafter RRG]{rrgmeth}. It is a ``passive'' method that measures the
Gaussian-weighted second order moments $I_{ij}=\frac{\sum w I x_i x_j }{\sum w I}$
of each galaxy  and corrects them using the Gaussian-weighted moments of the PSF
model.  RRG is well suited to the small, diffraction-limited PSF obtained from
space, because it corrects each moment individually, and only divides them to
form an ellipticity at the final stage. 

In an advance from previous implementations of KSB, and spurred by the findings of the Shear
TEsting Program \citep[STEP;][]{step2}, we allow the shear responsivity factor $G$ to vary as
a function of magnitude. The shear responsivity is the conversion factor between measured
galaxy ellipticity $e_i$ and the cosmologically interesting quantity shear $\gamma_i$. As
described in \citet{apjse_lea}, we have tested our pipeline on simulated images created with
the same \citet{shims} package used for STEP, but tailored specifically to the image
characteristics of the COSMOS data. We found it necessary to multiply our shears by a mean
calibration factor of $(0.86)^{-1}$, but then found the shear calibration $\langle m\rangle$
accurate to $0.3\%$, with a residual shear offset $\langle c\rangle=0.2\pm4\times10^{-4}$,
with no significant variation as a function of simulated galaxy size or flux. This is
particularly important in the measurement of a shear signal as a function of redshift. See
\citet{step1} or \citet{step2} for the definitions of the multiplicative $\langle m\rangle$
and additive $\langle c\rangle$ shear errors.

\subsection{Charge transfer effects}

As discussed further in \citet{apjse_rho}, the ACS WFC CCDs also suffer from
imperfect charge transfer efficiency (CTE) during readout. This causes flux to
be trailed behind objects, spuriously elongating them in a coherent direction
that mimics a lensing signal. Furthermore, since this effect is produced by a
fixed number of charge traps in the silicon substrate, it affects faint
sources (with a larger fraction of their flux being affected) more than bright
ones. Thus it is an insidious effect that also mimics an increase in shear
signal as a function of redshift. CTE trailing is a nonlinear transformation
of the image, and prevents traditional tests of a weak lensing analysis that
look at bright stars. As such, it is the most significant hurdle to overcome
in weak lensing analysis from space.

We are developing a method to remove CTE trailing at the pixel level.
Following the work of \citet{bristow} on STIS, this will push charge back to
where it belongs, as the very first stage in data reduction. Because an ACS
version of this algorithm is still under development, in this paper we correct
most of the CTE effect via a parametric model acting at the catalog level. We
assume that the spurious change in an object's apparent ellipticity
$\varepsilon$, is an additive amount that depends only upon the object's flux,
distance from the CCD readout register, and date of observation. In fact, we
also allowed variation with object size, although this had little effect. As
shown in \citet{apjse_rho}, this correction is sufficient for the full catalog
of more than 70 galaxies per arcmin$^2$ when considering mass reconstruction,
or circularly averaged statistics on small scales, where the signal is strong.
However, it is not adequate for the faintest galaxies, when considering
statistics on large scales, as we would like to do in this paper. Fortunately,
the galaxy flux level at which the CTE correction successfully removes the CTE
signal (leaving a residual signal one order of magnitude below the expected
cosmological signal),
appears to coincide with that for which reliable photometric
redshifts can be obtained for almost all objects.

\subsection{Photometric redshifts}
\label{sec:photozintro}

Reliable photometric redshift estimation is vital to the success of our 3D
shear measurement. For this reason, the COSMOS field has been observed from
the ground in a comprehensive range of wavelengths (Capak \etal\ 2006). Deep
imaging is currently available in the Subaru $B_J$, $V_J$, $g^+$, $r^+$,
$i^+$, $z^+$, $NB816$, CFHT $u^*$, $i^*$, CTIO/KPNO $K_s$, and SDSS
$u^\prime$, $g^\prime$, $r^\prime$, $i^\prime$, $z^\prime$ bands. The COSMOS
photometric redshift code was used as described in Mobasher \etal\ (2006).
This code contains a luminosity function prior in order to maximise the global
accuracy of photometric redshifts for the faintest and most distant
population. It returns both a best-fit redshift and a full redshift
probability distribution for each galaxy. The size of $68\%$ confidence limits
for each estimated redshift are well-modelled by $0.03(1+z)$ out to $z\sim
1.4$ and down to magnitude $I_{F814W}=24$ \citep{apjse_mob,apjse_lea}.

Before a large spectroscopic redshift sample becomes available to calibrate
the galaxy redshift distribution, our 3D analysis will be limited by the
reliability of photometric redshifts. We do
not impose a strict magnitude cut in the single $I_{F814W}$ band, but instead
using color information from many bands, and select those galaxies with
accurately measured redshifts. This includes $96\%$ of detected galaxies
brighter than $I_{F814W}=24$, and an incomplete sample fainter than that
\citep{apjse_lea}. The selection function, and the final redshift
distribution, thus depend upon the spectral energy distribution of individual
galaxies. However, since the background galaxies are unrelated to the
foreground mass that is lensing them, such incompleteness has no detrimental
effect on our analysis. 

We specifically select galaxies that are observed in the multi-color ground-based
data and that have a $68\%$ confidence limit in their redshift probability
distribution function smaller
than $\Delta z=0.5$. The latter cut primarily removes galaxies with double peaks in
the photometric redshift PDF due to redshift degeneracies. With the range of colors
currently observed in the COSMOS field, one particular degeneracy dominates: between
$0.1<z<0.3$ and $1.5<z<3.2$, where the 4000\AA\ break can be confused with coronal
line absorption features. At $z>1.5$, the 4000\AA\ break is well into the IR, where
sufficiently deep data are not yet available for conclusive identification.  To avoid
catastrophic errors between these specific redshifts, we therefore also exclude
galaxies with {\it any} finite probability below $z=0.4$ and above $z=1.0$. After
these cuts, we have redshift (and shear) measurements for 40 galaxies per arcmin$^2$.



\section{2D shear analysis} \label{sec:2dresults}

\subsection{2D source redshift distribution} \label{sec:2dzdist}

\begin{figure}[tb]
\plotone{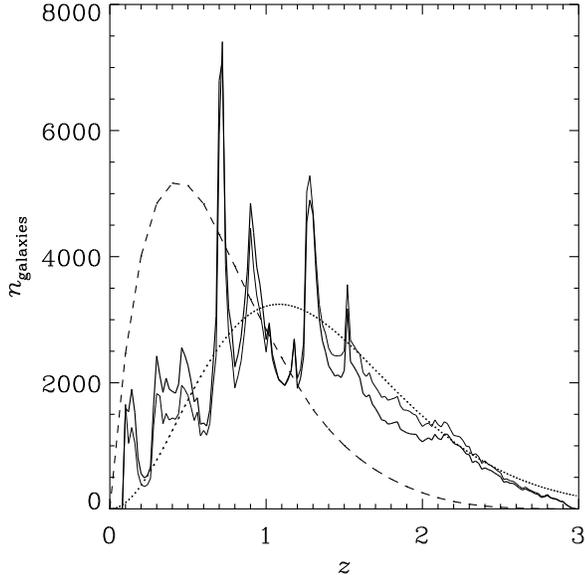}
\caption{The thin, solid line shows the distribution of the best-fit redshifts 
returned
by the COSMOS photometric redshift code (Mobasher \etal\ 2006) with a luminosity
function prior. The thick, solid line shows the distribution after accounting
for the different weights given to galaxies. In both cases, the bin size is
$\Delta z=0.02$. 
Peaks below $z\approx1.2$ correspond to real structures in the field, but the
artificial clustering at higher redshift is due to limitations in the finite
number of observed near-IR colors. 
The dashed curve shows the redshift
sensitivity function, assuming a $\Lambda$CDM universe with WMAP parameters.
The dotted line shows the redshift distribution that would have been expected,
with knowledge of only the median photometric redshift and a \citet{smailzdist}
fitting function.}
\label{fig:2dzdist}
\end{figure}

The distribution of galaxies with reliably measured shears \emph{and} redshifts
is shown in figure~\ref{fig:2dzdist}. The effects of cosmic variance are quite
apparent, with the spikes below $z\sim1.2$ all corresponding to known structures
in the field. Beyond that, the photometric redshifts are limited by the finite
number of observed colors for each galaxy, and the peaks at $z=1.3$, 1.5 and 2.2
arise artificially at locations where spectral features move between filters.
The median photometric redshift is $z_{\rm med}=1.26$.
To minimize the impact of galaxy shape measurement noise, we downweight the
contribution to the measured signal from faint and therefore noisier galaxies.
We apply a weight
\begin{equation}
w=\frac{1}{\sigma_\epsilon({\rm mag})+0.1} ~,
\end{equation}
\noindent where the rms dispersion of observed galaxy ellipticities is 
well-modelled by
\begin{equation}\label{eqn:sigmae}
\sigma_\epsilon({\rm mag})\approx 0.32+0.0014({\rm mag}-20)^3 ~.
\end{equation}
\noindent 
The error distribution of the shear estimators is discussed in more detail in
\citet{apjse_lea}.
After this weighting, the median photometric redshift is $z_{\rm
med}=1.11$. In most cosmic shear analyses to date, an estimate of this value is
all that was known about the redshift distribution. The smooth, dotted curve
shows the distribution that would have been obtained from a \citet{smailzdist}
fitting function

\begin{equation}
P(z)~\propto~z^\alpha~\exp{\left(-\left(1.41z/z_{\rm med}\right)^\beta\right)}
\label{eqn:smailzdist}
\end{equation}
\noindent with $\alpha=2$, $\beta=1.5$, $z_{\rm med}=1.26$ and an overall
normalization to ensure the correct projected number density of galaxies. This
would have been a better fit to the high redshift tail apparent in
figure~\ref{fig:2dzdist}, had the free parameter in the model, $z_{\rm med}$,
been $\sim1.17$.

Figure~\ref{fig:2dzdist} also shows the lensing sensitivity function
\begin{equation}
g(\chi) = 2 \int_{\chi}^{\chi_{h}} \eta(\chi')~
   \frac{D_A(\chi)D_A(\chi'-\chi)}{D_A(\chi')}~a^{-1}(\chi)~\rmd\chi'~,
\label{eq:theory:lensingweightfunction}
\end{equation}
\noindent of the observed source redshift distribution, where $\chi$ is a
distance in comoving coordinates (in which the power spectrum is measured),
$\chi_h$ is the distance to the horizon, $D_A$s are angular diameter distances,
(with the extra factor of $a^{-1}$ converting these into comoving coordinates),
and $\eta(\chi)$ is the distribution function of source galaxies in
redshift space, normalized so that
\begin{equation}
\int_0^{\chi_{h}} \eta(\chi) ~\rmd\chi = 1 ~.
\end{equation}
\noindent This represents the sensitivity of a projected lensing analysis to
mass overdensities, as a function of their redshift, and peaks at $z\sim0.4$,
about half-way to the peak of the source galaxy redshift distribution in terms
of angular diameter distance.

\subsection{2D shear correlation functions}

The 2D power spectrum of the projected shear field is given by
\begin{equation}
C^\gamma_{\ell} ~=~ \frac{9}{16} \left( \frac{H_{0}}{c} \right)^{4}
\Omega_{m}^{2}
  \int_{0}^{\chi_h} \left[ \frac{g(\chi)}{D_A(\chi)} \right]^{2}
  P\left(k, \chi\right)~\rmd\chi,
\label{eqn:shearpowerspectrum}
\end{equation}
\noindent where $\chi$ is a comoving distance; $\chi_h$ is the horizon distance;
$g(\chi)$ is the lensing weight
function; and $P(k,\chi)$ is the underlying 3D distribution of mass in the
universe. The two-point shear correlations functions can  be expressed
\citep{schneb} in terms of the projected power spectrum as
\begin{eqnarray}
C_1(\theta) & = &
\frac{1}{4\pi}\int_0^\infty C_\ell^\gamma ~\big[{\rm J}_0(\ell\theta)
+{\rm J}_4(\ell\theta)\big] ~\ell~\rmd\ell
\label{eqn:c1} \\
C_2(\theta) & = &
\frac{1}{4\pi}\int_0^\infty C_\ell^\gamma ~\big[{\rm J}_0(\ell\theta)
-{\rm J}_4(\ell\theta)\big] ~\ell~\rmd\ell ~.
\label{eqn:c2}
\end{eqnarray}

These can be measured by averaging over galaxy pairs, as
\begin{eqnarray}
C_{1}(\theta) & = &
  \big\langle ~ \gamma_{1}^r({\bf r}) ~
            \gamma_{1}^r({\bf r}+{\bf\theta}) ~
            \big\rangle
\label{eqn:cth_c1} \\
C_{2}(\theta) & = &
  \big\langle ~ \gamma_{2}^r({\bf r}) ~
            \gamma_{2}^r({\bf r}+{\bf\theta}) ~
            \big\rangle ~,
\label{eqn:cth_c2}
\end{eqnarray}
\noindent where $\theta$ is the separation between the galaxies and the
superscript~$^r$ denotes components of shear rotated so that $\hat\gamma_1^r$
($\hat\gamma_2^r$) in each galaxy points along (at 45$\degr$ from) the vector
between the pair. In practice, we compute this measurement in discrete bins of
varying angular scale. However, they will need to be integrated later, so to
keep this task manageable, we use fine bins of 0.1\arcsec throughout the
calculations, and only rebin for the sake of clarity in the final plots.

A third shear-shear correlation function can be formed,
\begin{equation}
C_{3}(\theta) ~=~
  \big\langle ~ \gamma_{1}^r({\bf r}) ~
            \gamma_{2}^r({\bf r}+{\bf \theta}) ~
            \big\rangle ~+~
  \big\langle ~ \gamma_{2}^r({\bf r}) ~
            \gamma_{1}^r({\bf r}+{\bf \theta}) ~
            \big\rangle ~,
\label{eqn:cth_c3}
\end{equation}
\noindent for which parity invariance of the universe requires a zero
signal. The presence or absence of $C_3(\theta)$ can therefore be used as a
first test for the presence of systematic errors in our measurement, although
many systematics can still be imagined that would not show up in this test.

\begin{figure}[tb]
\plotone{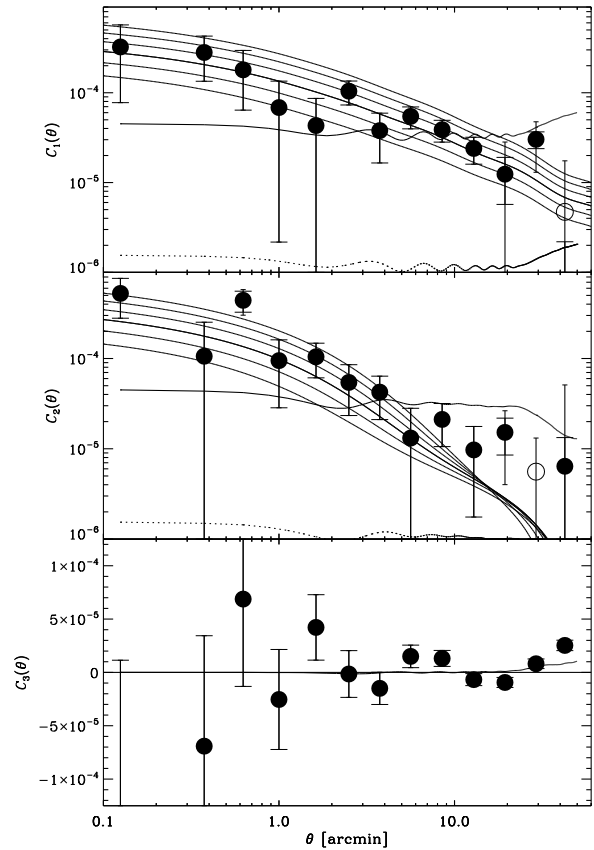}
\caption{Correlation functions of the 2D shear field.
The open circles indicate negative values. The inner error bars show statistical
errors only; the outer error bars, visible only on large scales, 
also include the contribution of cosmic variance.
The six parallel curves show theoretical
predictions for a flat $\Lambda$CDM cosmology with $\Omega_m=0.3$ and $\sigma_8$ varying from 0.7
(bottom) to 1.2 (top). The roughly horizontal lines indicate the
level of the spurious signal due to CTE trailing before and after 
correction.}
\label{fig:2dcth}
\end{figure}

The 2D shear correlation functions measured from the entire COSMOS survey are shown
in figure~\ref{fig:2dcth}. Note that the measurements on scales smaller than
$\sim1\arcmin$ are new. For a given survey size, these are obtained more easily from
space than from the ground because of the higher number density of resolved galaxies.

The additional, spurious signal that would have been obtained without
correction for CTE trailing is shown as roughly horizontal, solid lines in
figure~\ref{fig:2dcth}. This was calculated by recomputing the correlation
functions but, rather than constructing a shear catalog by subtracting the CTE
contamination from each galaxy's raw shear measurement, the CTE contamination
was used as a direct replacement. An estimate of the residual CTE
contamination for the galaxy population {\it after} correction, according to
the performance evaluation in \citet{apjse_rho}, is shown as dotted lines.
Although this is now below the signal, the uncorrected level was more than an
order of magnitude larger than the signal on large scales. Minimizing CTE by
careful hardware design to avoid the need for this level of correction will be
a vital aspect of dedicated space-based weak-lensing missions in the future.

\subsection{Error estimation and verification}
\label{sec:error}

The error bars in figure~\ref{fig:2dcth} include both statistical errors due to intrinsic
galaxy shape noise within the survey, and the effect of sample (``cosmic'') variance due to
the finite survey size. The shape noise dominates on small angular scales, and the cosmic
variance on scales larger than $\sim10\arcmin$. Surveys covering a similar area but in
multiple lines of sight, such as ACS parallel data \citep[][Rhodes \etal\ in
preparation]{gemscs2}, will suffer less from the latter effect. 

The statistical shape noise is easy to measure from the gaalxy population. To measure the
sample variance, we split the COSMOS field into four equally-sized quadrants and recalculate
the correlation functions in each. Of course, large-scale correlations in the mass
distribution mean that the four adjacent quadrants are not completely independent at large
scales, and the measured variance underestimates the true error. To correct for this effect,
we artificially increased the measured errors on 20--40$\arcmin$ scales by $15\%$, in line
with initial calculations.

\begin{figure}[tb]
\plotone{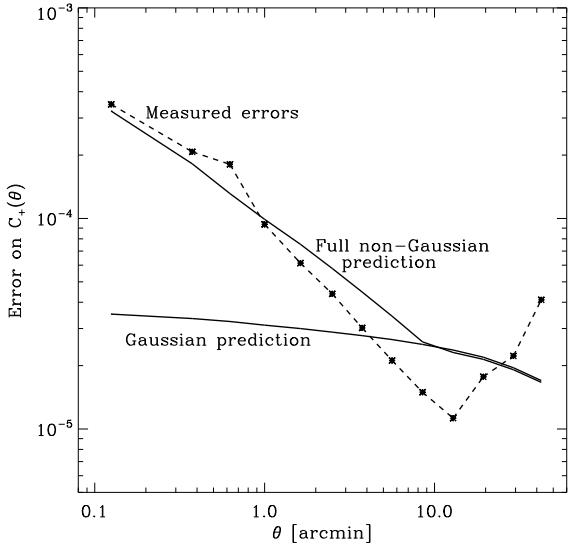}
\caption{Comparison of the error bars that we measured from the data, to advance 
predictions from
\citet{semvariance}, obtained by raytracing through $n$-body simulations of large-scale
structure. The two solid lines show the predictions assuming a Gaussianised mass
distribution (bottom) and with the full, non-Gaussian distribution (top).}
\label{fig:2derrorssim}
\end{figure}

After the fact, we have compared our final error bars to independent predictions from a full
raytracing analysis through $n$-body simulations by \citet{semvariance}.
Figure~\ref{fig:2derrorssim} shows the predicted and observed $1\sigma$ errors on
$C_+(\theta)\equiv C_1(\theta)+C_2(\theta)$ (assuming 40 background galaxies per square
arcminute in the simulations, distributed in redshift with $z_{\rm med}=1.11$, and with
$\sigma_\varepsilon=0.32$). Averaging across all thirteen angular bins with equal weight, the
mean ratio between our measured error and the predicted non-Gaussian error is 0.994. Future
work may therefore improve the error estimation, but in the COSMOS field at least, our
quadrant technique reaches a level of precision sufficient for this paper.


\begin{figure}[tb]
\plotone{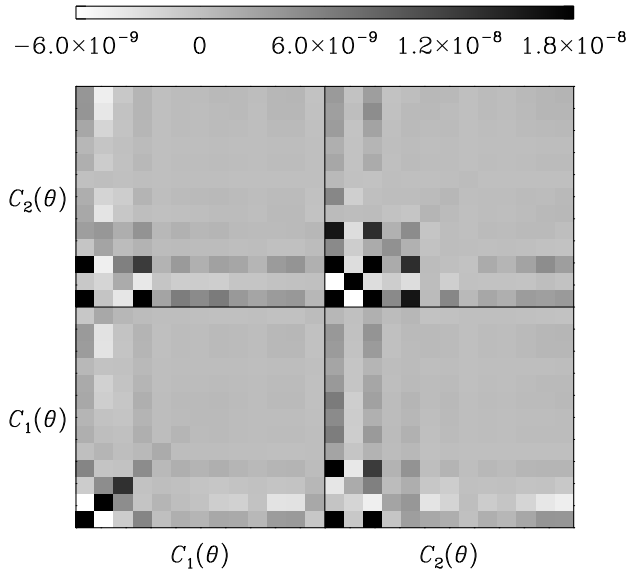}
\caption{Covariance matrix for the 2D correlation functions $C_1(\theta)$ and
$C_2(\theta)$ shown in
figure~\ref{fig:2dcth}, obtained by splitting the COSMOS field into four
quadrants and performing the analysis separately in each.
The diagonal elements illustrate the size of
the errors in each of the thirteen $\theta$ bins, and the off-diagonal
elements illustrate how much the measurements are correlated.
The color scale is logarithmic.}
\label{fig:2dcovariance}
\end{figure}

We also use the quadrant technique to measure the full covariance matrix between each angular
bin. As shown in figure~\ref{fig:2dcovariance}, the off-diagonal elements are non-zero. This
is expected even in an ideal case, because the same source population of galaxies is used to
construct pairs separated by different amounts. Nor are the upper-left and lower-right
quadrants of figure~\ref{fig:2dcovariance} expected to be zero: the same pairs go into the
calculation of both $C_1(\theta)$ and $C_2(\theta)$; and after deconvolution from the PSF,
$\hat{\gamma}^r_1$ and $\hat{\gamma}^r_2$ are no longer formally independent. We will use the
full, non-diagonal covariance matrix during our measurement of cosmological parameters in
\S\ref{sec:constraints}.

The final datum in the $C_3(\theta)$ panel of figure~\ref{fig:2dcth} is significantly ($\sim5\sigma$)
non-zero. This may be real: a finite region may {\it not} be parity invariant on scales comparable to
the field size. But even if this does indicate a systematic problem, it is not as troubling as it
appears, because on this scale the error bars are large for $C_1(\theta)$ and $C_2(\theta)$, so the
point carries very little weight. For a possible explanation, note that the spurious $C_3$ signal has
the same sign as the uncorrected CTE signal. On scales that span almost the entire COSMOS survey, one
of the galaxies in a pair must lie near the edge of the survey field -- which was observed last, and
suffers most from CTE degradation. If the temporal dependence of the CTE signal is not linear, as we
have assumed, the spiral observing strategy to cover the field would produce a similar CTE pattern
(and a coherent residual signal) in all four quadrants. This could create an additional
$C_3(40\arcmin)$ signal, with error bars underestimated by our quadrant method. Resolving this issue
requires CTE data from a longer time span, or more data separated by 20--$40\arcmin$. Such analysis
may be feasible with ACS parallel data (Rhodes \etal\ in preparation) but is not possible here.

\subsection{2D shear variance}

\begin{figure}[tb]
\plotone{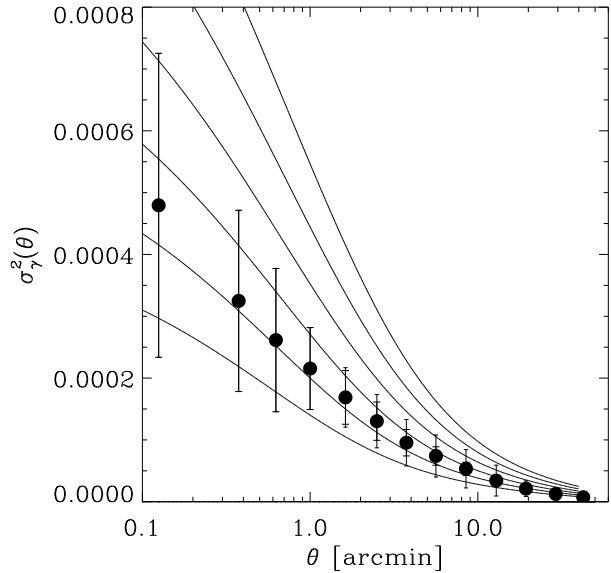}
\caption{Variance of the 2D shear signal in circular cells of varying size.
Solid lines show predictions in a concordance cosmology with $\sigma_8$ varying
as in figure~\ref{fig:2dcth}. Note that adjacent data points are highly correlated.}
\label{fig:2dpth}
\end{figure}

For historical reasons, cosmic shear results are often expressed as the
variance of the shear field in circular cells on the sky. For a top-hat cell
of radius $\theta$, this measure is related to the shear correlation
functions by
\begin{equation}
\sigma_\gamma^2 \equiv \langle ~ |\overline{\gamma}|^2 \rangle
                \approx \frac{2}{\theta^2}
\int_0^\theta \Big[ C_1(\vartheta)+C_2(\vartheta) \Big]~
     \rmd\vartheta ~,
\end{equation}
\noindent where we have used a small angle approximation. 
Note that the signal is more strongly correlated on different angular scales
in this form than it is when expressed as correlation functions. The results
are shown in figure~\ref{fig:2dpth}.

\subsection{2D E-B decomposition}

The correlation functions can also be recast in terms of non-local $E$
(gradient) and $B$ (curl) patterns in the shear field \citep{crit00,peneb}.
Gravitational lensing is expected to produce only $E$ modes, except for a very
low level of $B$ modes due to lens-lens coupling along a line of sight
\citep{schneb}. It is commonly assumed that systematic effects would affect
both $E$- and $B$-modes equally. The presence of a non-zero $B$ mode is
therefore a useful indication of contamination from other sources.

\begin{figure}[tb]
\plotone{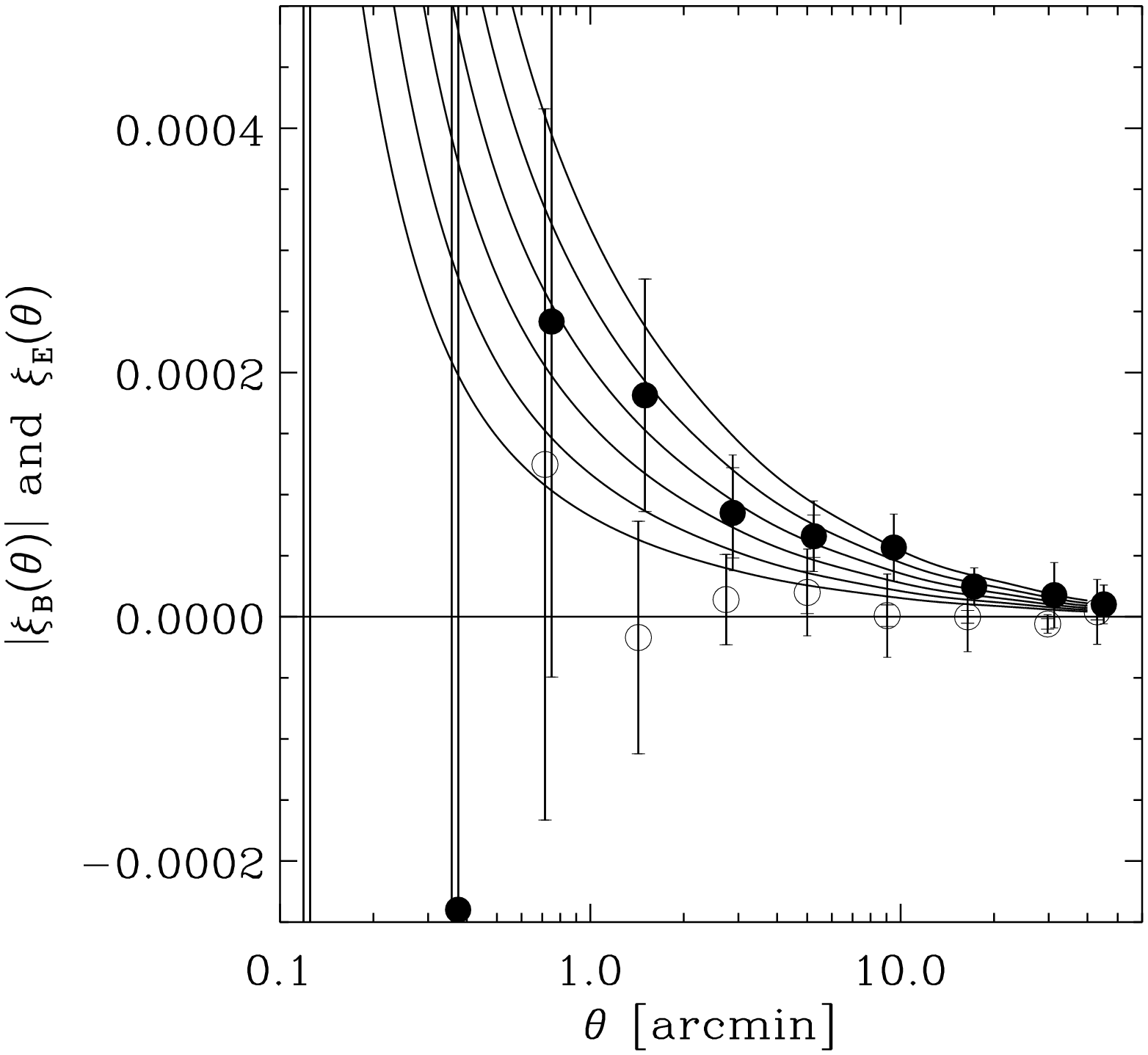}
\plotone{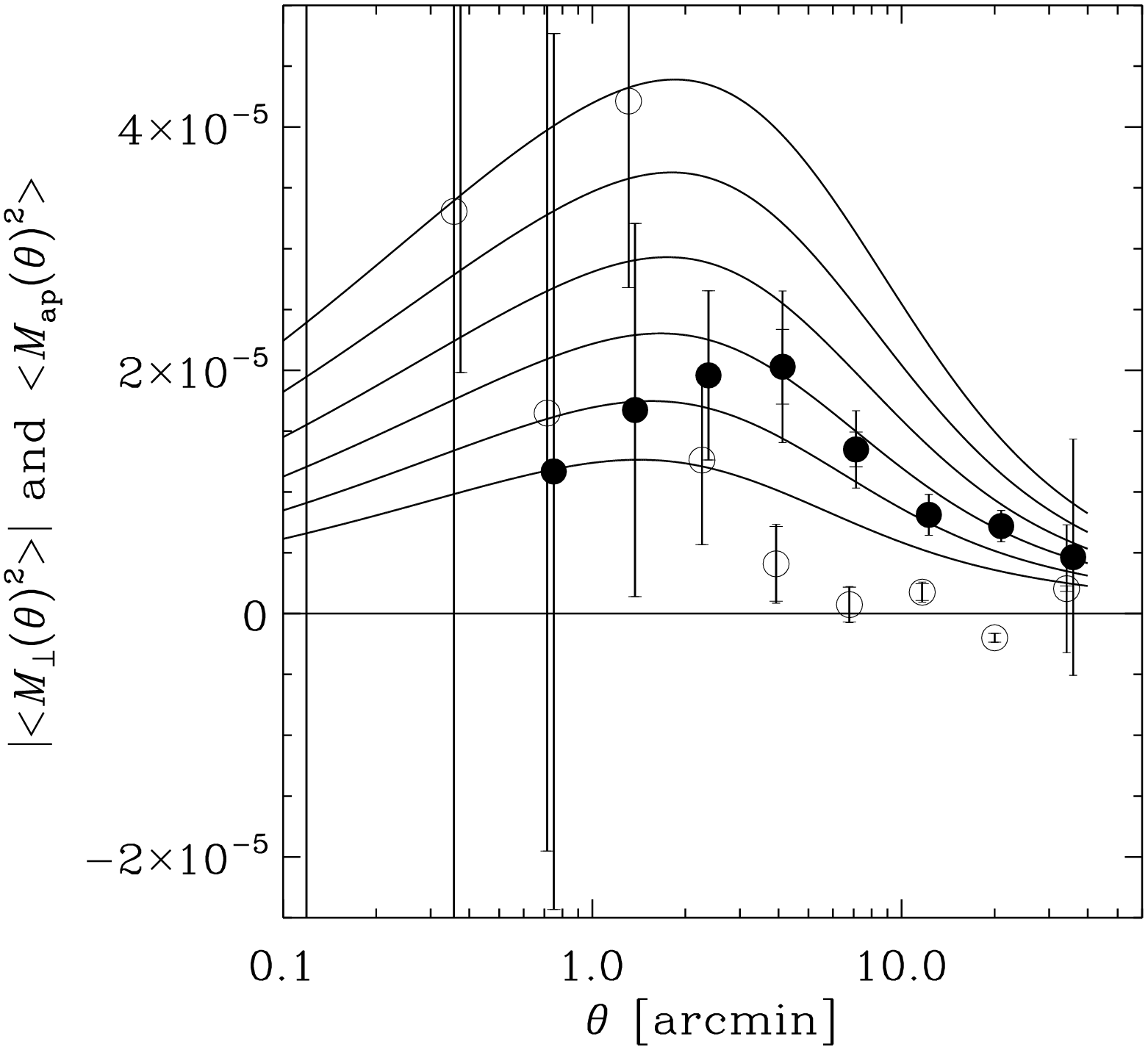}
\caption{$E$-$B$ decompositions of the 2D shear field. The top panel shows the
statistics that formally require an integral over the measured correlation
functions to infinite scales, and the bottom panel shows those that formally
require an integral from zero. Solid points show the $E$-mode, and open points
show the $B$-mode. The lines show predictions in a concordance cosmology with
$\sigma_8$ varying as in figure~\ref{fig:2dcth}. Note that adjacent data points are
in the top panel are highly correlated.}
\label{fig:2deb}
\end{figure}

$E$- and $B$-modes correspond to patterns within an extended region on the sky,
and cannot be separated locally. As a result, this operation formally requires
an integration of the shear correlation functions over a wide range of angular
scales. Two mathematical functions have been developed \citep{crit00,schneb},
which each include an integral over only small scales or large scales (see
\citet{ringtest} for a new suggestion to construct a third). However, neither
integral is ideal in practice, because our correlation functions are only well
measured on scales between $\sim 0.5\arcmin$ and $40\arcmin$. The absence of
complete data introduces an unknown constant of integration, and it is not
possible to uniquely split this measured shear field into distinct $E$- and
$B$-mode components. As a practical attempt to estimate this constant, we
extrapolate data into the unknown r\'egime, using predictions from the best-fit
cosmology that is determined in \S\ref{sec:constraints}.

The signal on large angular scales is small, and the corresponding integrals
require the least correction. To calculate these, we first define $C_{+} \equiv
C_1 + C_2$ and $C_-\equiv C_{1}-C_{2}$. Then we can compute
\begin{equation}
\xi_E(\theta) \equiv
           C_1(\theta)+2\int_\theta^\infty
           \left( 1-\frac{3\theta^2}{\vartheta^2} \right)
           \frac{C_{-}(\vartheta)
           }{\vartheta}~\rmd\vartheta ~,
\end{equation}
which contains only the $E$-mode signal and
\begin{equation}
\xi_B(\theta) \equiv
           C_2(\theta)-2\int_\theta^\infty
           \left( 1-\frac{3\theta^2}{\vartheta^2} \right)
           \frac{C_{-}(\vartheta)}{\vartheta}~\rmd\vartheta ~,
\label{eqn:cth_eb}
\end{equation}
which contains only the $B$-mode signal.
It is generally necessary to add a function of $\theta$ (not
only a constant of integration) to $\xi_E(\theta)$ and subtract it from
$\xi_B(\theta)$ \citep[c.f.][]{peneb}.

The components can also be separated via the variance of the aperture mass
statistic $\Map(\theta)$. This is obtained from a weighted mean of the
tangential ($\gamma_t$) and radial ($\gamma_r$) components of shear relative to
the center of a circular aperture. This statistic is given by
\begin{equation}
M_{\rm ap}(\theta) \equiv \int_0^\infty
~W\left(|\vec\vartheta|;\theta\right)~\gamma_t\left(\vec\vartheta\right)
~{\rm d}^2\vec\vartheta
\end{equation}
which contains only contributions from the $E$-mode signal and
\begin{equation}
M_\perp(\theta) \equiv \int_0^\infty
~W\left(|\vec\vartheta|;\theta\right)~\gamma_r\left(\vec\vartheta\right)
~{\rm d}^2\vec\vartheta~,
\end{equation}
which contains only the $B$-mode signal, where
$W(|\vec\vartheta|)$ is a compensated filter.  
We adopt a compensated ``Mexican hat'' weight function
\begin{equation}
W(\vartheta;\theta) = \frac{6}{\pi\theta^2} \frac{\vartheta^2}{\theta^2}
\left(1-\frac{\vartheta^2}{\theta^2}\right) {\rm
H}(\theta-\vartheta) ~,
\end{equation}
where $\theta$ defines an angular scale of the aperture and the
Heaviside step function ${\rm H}$ truncates the weight function on large scales.

\citet{schneb} derived
expressions for the variance of these statistics as the aperture is moved across
the sky. These reqire integrals over the correlation functions
from small scales
\begin{eqnarray}
\left\langle M^2_{\rm ap}\right\rangle(\theta) \equiv
\frac{1}{2}\int_0^{2\theta}\frac{\vartheta}{\theta^2} \Bigg[
 C_+(\vartheta)T_+\left(\frac{\vartheta}{\theta}\right) ~~~~~~~~ \nonumber \\
 +~
 C_-(\vartheta)T_-\left(\frac{\vartheta}{\theta}\right)
\Bigg]~{\rm d}\vartheta \label{eqn:MapVar}~ \\
\left\langle M^2_\perp\right\rangle(\theta) \equiv
\frac{1}{2}\int_0^{2\theta}\frac{\vartheta}{\theta^2} \Bigg[
 C_+(\vartheta)T_+\left(\frac{\vartheta}{\theta}\right) ~~~~~~~~ \nonumber \\ 
 -~
 C_-(\vartheta)T_-\left(\frac{\vartheta}{\theta}\right)
\Bigg]~{\rm d}\vartheta \label{eqn:MtangVar},
\end{eqnarray}
where
\begin{eqnarray}
T_+(x) ~=~
\frac{6(2-15x^2)}{5}\left[1-\frac{2}{\pi}\arcsin\left(\frac{x}{2}\right) \right]
~+~~~~~\\
        \frac{x\sqrt{4-x^2}}{100\pi} \big(120 + 2320x^2 - 754x^4 + 132x^6 - 9x^8\big) \nonumber
\end{eqnarray}
\begin{eqnarray}
T_-(x) ~=~ \frac{192}{35\pi}x^3\left(1-\frac{x^2}{4}\right)^{7/2}~~~~~~~~~~~~~~~~~~~~
\end{eqnarray}
for $x<2$ and $T_+(x)=T_-(x)=0$ for $x\geq 2$.
We again estimate the constant of integration by extrapolating our data with
theoretical predictions in cosmological model preferred by the rest of the data.

From figure~\ref{fig:2deb}, we can see that $\xi_B(\theta)$ is consistent with
zero on all scales. The noise is particularly large on small scales, and  the
rather unstable $M^2_\perp(\theta)$ is affected on scales up to $\sim1\arcmin$
by the first bin.

\section{3D shear analysis} \label{sec:3dresults}

\begin{figure}[tb]
\plotone{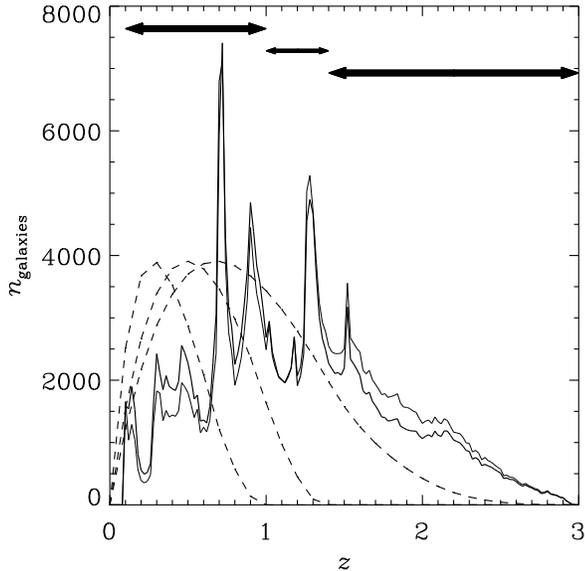}
\caption{The thin, solid line shows the redshift distribution of 
source galaxies. The thick, solid line shows their distribution after accounting
for the magnitude-dependent weighting scheme. In both cases, the bin size is
$\Delta z=0.02$.
The dashed lines show (artificially normalized) redshift sensitivity
curves obtained by slicing this distribution into the discrete redshift bins
indicated by the arrows at the top.}
\label{fig:3dzdist}
\end{figure}

\subsection{Correlation function tomography} \label{sec:tomography}

We now split the catalog into three discrete redshift bins and, as before,
calculate the correlation functions using all pairs of galaxies within each bin.
The redshift bins are chosen in consideration of the particular color
information available. Degeneracies in the photometric redshift estimation cause
galaxies with a flat distribution in redshift to cluster artificially around
$z=1.3$, 1.6 and 2.2. An excess at these positions is evident in
figure~\ref{fig:2dzdist}. We therefore pick bins with boundaries away from these
values, and with widths similar to the size of the local peaks in the resdshift
distribution.  For COSMOS, suitable bins are $0.1\leqslant z \leqslant 1$, $1< z
\leqslant 1.4$ and $1.4 < z \leqslant 3$. This scheme conveniently divides up
the galaxies fairly evenly, with the slices each containing $32\%$, $24\%$ and
$44\%$ of the galaxies. Unfortunately, the last bin can not be further
sub-divided without deeper IR or UV data. The redshift slices and their
resulting lensing sensitivity functions are illustrated in
figure~\ref{fig:3dzdist}.

\begin{figure}[tb]
\plotone{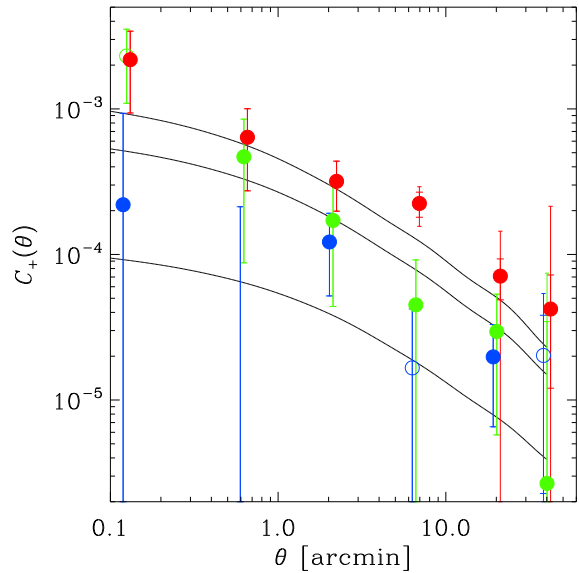}
\caption{Evolution of the cosmic shear two-point correlation function signal
with increasing redshift.
The series of data points, from bottom to top, show measurements from
slices between redshifts 0.1, 1, 1.4, and 3. The black curves
show predictions from a flat $\Lambda$CDM model with $\Omega_{\rm m}=0.3$ and
$\sigma_8=0.85$, for the same slices, increasing in redshift from the bottom
to the top. Open circles depict negative values.
\label{fig:3dcth}}
\end{figure}


Figure~\ref{fig:3dcth} shows the increasing two point correlation function signal for
pairs of source galaxies as a function of redshift, where both galaxies are in the
same redshift bin. Since the measurements in the redshift bins are much more noisy
than those from the projected 2D analysis, we plot $C_+(\theta)\equiv
C_1(\theta)+C_2(\theta)$ in figure~\ref{fig:3dcth}. Theoretical predictions for the
correlation functions are obtained for each slice by replacing the lensing weight
function $g(z)$ in equation~(\ref{eqn:shearpowerspectrum}) by those shown in
figure~\ref{fig:3dzdist}, and obtained from only the galaxies in a given slice.
Because the effective lensing volume $\int g(z)~{\rm d}z$ increases for succesive
redshift bins, the signal increases with $z$.

\begin{figure}[tb]
\plotone{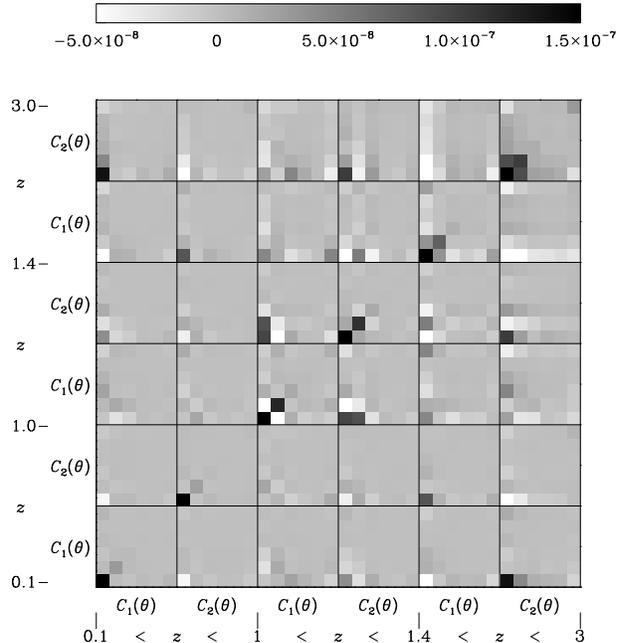}
\caption{Covariance matrix of correlation function data used in the 3D cosmic
shear analysis. Note that this includes only the three correlation functions
where both galaxies are in the same redshift bin. An additional three
correlation functions can be formed from pairs in which the galaxies come from
different slices, but these are not shown in this plot for the sake of clarity.}
\label{fig:3dcovariance}
\end{figure}

Figure~\ref{fig:3dcovariance} shows the measured covariance matrix for the 3D
correlation functions. The degree of correlation between the lowest and highest
redshift bins, primarily evident on small scales, is unexpected. Had it been
significant on all scales, a likely explanation would have been cross-contamination
of the bins by galaxies from other redshifts (the well-known degeneracy between low
and high redshift from photo-$z$ estimation is discussed in \S\ref{sec:photozintro}).
Had the covariance been equally evident in all three bins, likely explanations could
have been: interference of intrinsic alignments like those suggested by
\citep{intalinterfere}; and imperfect correction for PSF variation or DRIZZLE-related
pixellisation effects unaccounted for on small scales. In practice, the most likely
explanation is a combination of several such effects, each at a low level.

\begin{figure}[tb]
\plotone{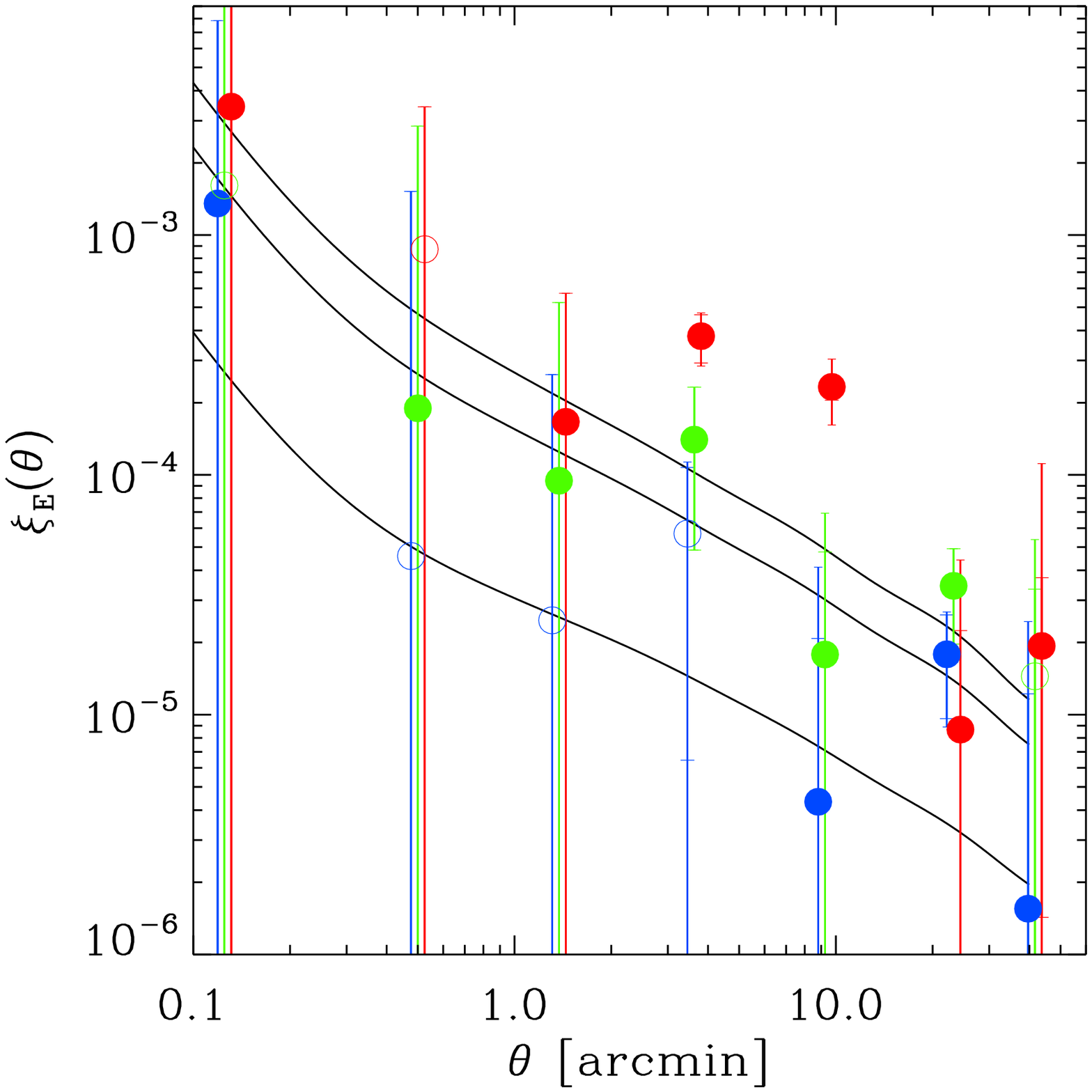}
\plotone{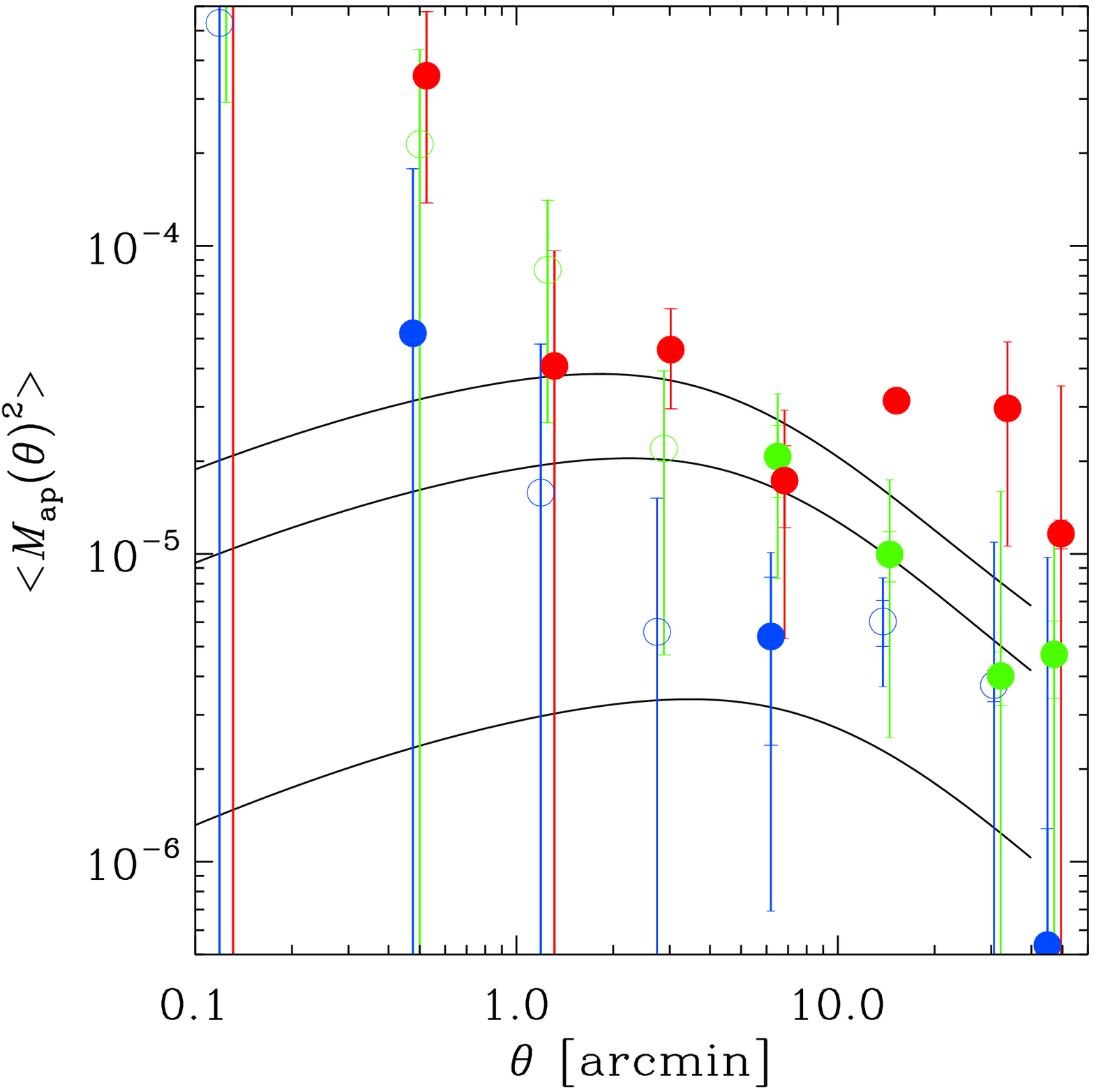}
\caption{$E$-$B$ decomposition of the 3D cosmic shear signal, in different
redshift bins, colored as in figure~\ref{fig:3dcth}. For clarity, only the 
$E$-modes are shown. Open circles depict negative values.
The $B$-modes are as noisy, but are consistent with zero.
Note that adjacent data points are highly correlated.}
\label{fig:3deb}
\end{figure}

Although the signal in the individual slices is noisy, we have attempted an
$E$-$B$ decomposition in figure~\ref{fig:3deb}, using the same two
statistics as were applied to the 2D analysis. The integrals over the noisy
correlation functions are particularly ill-defined at $\theta<1\arcmin$.
Nevertheless, the signal increases to high redshift, matching the theoretical 
expectation for this measurement.

\subsection{Growth of structure} \label{sec:growth}

The total $E$-mode signal corresponds to the integrated mass density along a
line of sight, weighted by the lensing sensitivity function. The evolving
$E$-mode signal in figures~\ref{fig:3dcth} and \ref{fig:3deb} grows towards
high redshift due to the increasing volume that it probes, and in which mass
structures are located. This offers constraints on the large-scale geometry of
the universe. But, if we are more interested in the mass structures
themselves, this function of $\theta$ in fixed redshift slices can be recast
into a function of $z$ for fixed angular scales. We shall now suggest a new
way of viewing this data, which stays close to measurable quantities, but
offers a new insight into the underlying structure formation.

Each data point in figure~\ref{fig:3dcth} corresponds to the amount of mass
within an effective volume. This volume is described in azimuthal directions
by Bessel functions, and in the redshift direction by the lensing sensitivity
function $g(z)$. Assuming the best-fit cosmology from \S\ref{sec:constraints}
to fix the geometry of the universe, we can divide by this volume, and obtain
a quantity proportional to mass density. In practice, to increase the
signal-to-noise of a measurement that will involve many redshift bins, we do
not restrict the measurement to only those pairs within a given redshift
slice, as before. We require the nearer galaxy to be inside the slice, but
then compute correlation functions using all galaxies behind it. The more
distant galaxy has then been lensed by anything the foreground has been lensed
by. The effect is merely to change the (squared) lensing sensitivity function
to the product of the sensitivity function for the slice galaxies with that of
the background distribution. This creates a new, effective $g(z)$ that peaks
at slightly higher redshift, but is still zero behind the nearest galaxy.

\begin{figure}[tb]
\plotone{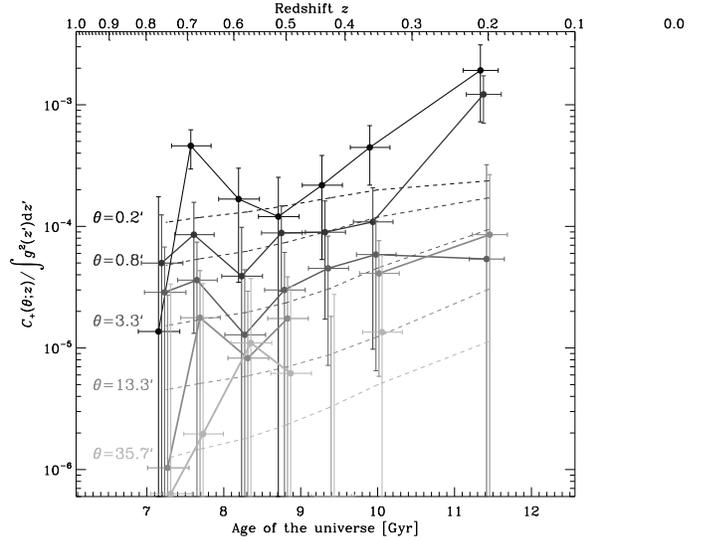}
\caption{The growth of structure over cosmic time. This links the cosmic shear
signal on fixed angular scales as a function of redshift (rather than the
other way round, as in previous figures). Data points are located at the peak of
the lensing sensitivity function for each set of source galaxies. The source 
galaxies themselves are approximately twice as far away.
The different colors distinguish different angular scales. 
For each of these, the dashed line shows the theoretical expectation, assuming 
the best-fit cosmological model from \S\ref{sec:constraints}.}
\label{fig:3dgos}
\end{figure}

Figure~\ref{fig:3dgos} thus shows
\begin{eqnarray}
G(z;\theta) 
 &\equiv& C_+(\theta;z)~\Big/\int_0^z g^2(z^\prime)~{\rm d} z^\prime \\
 &   =  & \frac{1}{2\pi}\frac{\int C_\ell^\gamma(\theta;z)~
J_0(\ell\theta)~\ell{\rm d}\ell}{\int_0^z g^2(z^\prime)~{\rm d} z^\prime}~,
\end{eqnarray}
\noindent the growth of power on different angular scales. The foreground mass
is most likely to lie near the peak of the sensitivity function, so we place
the data points at this redshift. In practice, it could lie anywhere
within $g(z)$, so we overlay error bars in $z$ equal to the rms of $g$
about its peak. Theoretical predictions of this quantity are overlaid,
assuming a flat, $\Lambda$CDM cosmology, with the best-fit parameters found in
\S\ref{sec:constraints}. 

The growth towards $z=0$ represents a combination of the physical growth of
structure and the mixing of fixed physical scales at different redshifts into a
measurement at one apparent angular scale. Both of these effects act in the same
sense, to increase the signal towards the present day. This is in contrast to
the cosmic shear signal in figures~\ref{fig:3dcth} and \ref{fig:3deb}, which
itself increases towards high redshift. On large scales, the small cosmic shear
signal makes the measurement fairly noisy. On intermediate scales, the data
closely follow the predictions. The lowest redshift point is obtained from pairs
of galaxies where the nearest is between $z=0.1$ and $z=0.7$. We speculate that
the apparently significant upturn at low $z$ and on small scales might be caused
by contamination of that redshift bin by high redshift galaxies. These could
have been caught by the photometric redshift degeneracy discussed in
\S\ref{sec:photozintro} and would contain a apparently spurious signal when
moved to low redshift. The accuracy of the photometric redshifts may therefore
be limiting the precision of this measurement.

\section{Constraints on cosmological parameters} \label{sec:constraints}

\subsection{2D parameter constraints} \label{sec:2dconstranits}

We now use a Maximum Likelihood method to determine the constraints set by our
2D observations of $C_1(\theta)$ and $C_2(\theta)$ upon the cosmological
parameters $\Omega_{\rm m}$, the total mass-density of the universe, and
$\sigma_8$, the normalization of the matter power spectrum at 8 $h^{-1}$ Mpc.
We assume a flat universe, with a Hubble parameter $h=0.7$.

We closely follow the approach of \citet{xwht}, obtaining theoretical predictions for
the linear transfer function from the fitting functions of \citet{bbks} and for the
non-linear power spectrum using the fitting functions of \citet{smith}. The
theoretical correlation functions are first calculated from
equation~(\ref{eqn:shearpowerspectrum}) in a three dimensional grid spanning
variations in $\Omega_{\rm m}$ from 0.05 to 1.1, $\sigma_8$  from 0.35 to 1.4 and the
power spectrum shape parameter $\Gamma$ from 0.13 to 0.33. We used the full redshift
distribution of source galaxies (after correction for weighting) shown in
figure~\ref{fig:2dzdist}.

\begin{figure}[tb]
\vspace*{5mm}
\epsscale{0.84}
\plotone{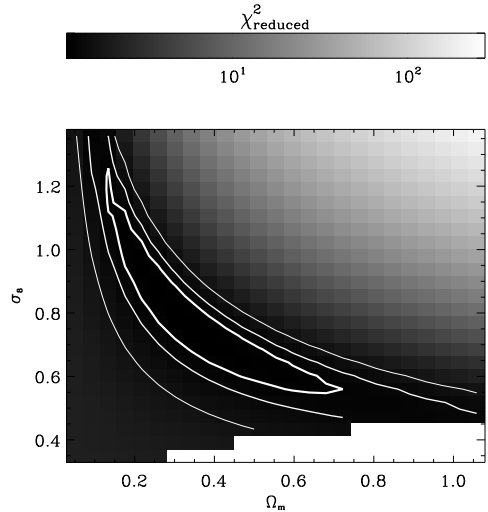}
\epsscale{1}
\caption{Constraints on cosmological parameters 
from a traditional 2D cosmic shear analysis, after marginalization over other 
free parameters.
In order of decreasing thickness, the contours indicate $68.3\%$, $95.4\%$ and
$99.7\%$ confidence limits due to statistical errors; additional uncertainty
potentially contributed by sources of systematic error are discussed in the
text. The grey scale background is logarithmic, and shows $\chi^2$ divided by
the number of degrees of freedom in the data. The white area at the bottom-right
was excluded because the \citet{smith} fitting functions could not be evaluated
without unreasonable extrapolation of the non-linear matter power spectrum to
physical scales smaller than $0.1h^{-1}$kpc. This can be compared to the much
tighter constraints from the full 3D analysis in figure \ref{fig:3ds8om}.}
\label{fig:2ds8om}
\end{figure}

We then fitted the observed shear correlation functions 
$\vec{d}(\theta)=\{C_1(\theta),C_2(\theta)\}$ to
the theoretical predictions calculated at the centers of each bin
$\vec{t}(\theta)$, computing the log-likelihood function
\begin{eqnarray}
\chi^2(\vartheta,\Omega_{\rm m},\sigma_8,\Gamma) = \big(\vec{d}(\theta) - \vec{t}(\theta,\Omega_{\rm m},\sigma_8,\Gamma) \big)^T
~ {\rm cov}(d)^{-1}
~ \big(\vec{d}(\vartheta) - \vec{t}(\vartheta,\Omega_{\rm m},\sigma_8,\Gamma) \big)
\nonumber
\label{eqn:chisq}
\end{eqnarray}
\noindent throughout the grid, where ${\rm cov}(d)$ is the covariance matrix in
figure~\ref{fig:2dcovariance}. In an advance of earlier incarnations, we perform
the matrix inversion via a singular value decomposition (SVD), and discard all
eigenvalues not within machine precision of the largest. We do not include the 
multiplicative factor suggested by \citet{cinverse}. We then marginalize over
$\Gamma$ with a Gaussian prior centered on 0.19 and with an rms width of 15\%
\citep{2df01}. To
compute confidence contours, we numerically integrate the likelihood function
\begin{equation}
L(\Omega_{\rm m},\sigma_8) ~=~ e^{-\chi^2/2} ~.
\end{equation}

Our constraints on cosmological parameters from this 2D analysis are presented as a
projection through parameter space in figure~\ref{fig:2ds8om}. The contours represent
statistical errors, including full non-Gaussian sample variance. Formally, the
best-fit model has $\Omega_{m}=0.30$, $\sigma_{8}=0.81$ and $\Gamma=0.21$, and this
achieves $\chi^2_{\rm reduced}\equiv\chi^2/(n_{\rm param}-3)=1.10$ in 23 degrees of
freedom. However, there is a well-known degeneracy between $\Omega_{\rm m}$ and
$\sigma_8$ when using only two-point statistics. Changing the $\Gamma$ parameter
slides the contours back and forth along this valley, and marginalization over this
parameter also slightly increases the minimum $\chi^2$. After marginalization, a good fit to our 68.3\%
confidence level from statistical errors is given by
\begin{equation}
\sigma_{8} \left( \frac{\Omega_{m}}{0.3} \right)^{0.44} = 0.81 \pm 0.075 ~,
\label{eqn:s8om}
\end{equation}
with $0.15\leqslant\Omega_{\rm m}\leqslant 0.7$.

\citet{xwht} were unable to use the full covariance matrix due to
instabilities in the matrix inversion, so had set to zero any elements in the 
covariance of $C_1(\theta)$ with $C_2(\theta)$ (these are the bottom-left and
top-right quarters in figure~\ref{fig:2dcovariance}). This problem has been
resolved in the present work by the use of an SVD. 
However, if we discard half of the covariance
matrix as in \citet{xwht}, we obtain parameter constraints
\begin{equation}
\sigma_{8} \left( \frac{\Omega_{m}}{0.3} \right)^{0.44} = 0.83 \pm 0.07 ~.
\end{equation}
\noindent If we discard {\it all} of the off-diagonal elements in the covariance matrix, we obtain
\begin{equation}
\sigma_{8} \left( \frac{\Omega_{m}}{0.3} \right)^{0.44} = 0.84 \pm 0.065 ~.
\end{equation}
\noindent The slightly smaller error bars are expected, but the shift in the
best-fit value relative to result~\eqref{eqn:s8om} is not. This effect might go some way towards explaining the
higher than usual value obtained for this quantity in \citet{xwht}.

Note that all of the above constraints incorporate only {\it statistical} 
sources of error; although these do include
non-Gaussian sample variance and marginalization over other parameters. 
We can propagate the various sources of potential
{\it systematic} error by noting that
\begin{equation}
C_i(5')\propto
\Omega_{\rm m}^{1.46}\sigma_8^{2.45}z_s^{1.65}\Gamma^{-0.11}(P^\gamma)^{-2},
\label{eqn:coszgp}
\end{equation}
\noindent for $i\in\{1,2\}$ in a fiducial $\Lambda$CDM cosmological model with
$\Omega_{\rm m}=0.3$, $\Omega_\Lambda$=0.7, $\Gamma=0.21$ and $\sigma_8=1.0$. 
Adding an uncertainty equivalent to
10\% in the median source redshift, a 6\% shear calibration
uncertainty \citep[see][]{apjse_lea,step1,step2}, and an empirically estimated
binning instability \citep[\cf][]{xwht} to our constraint from the full
covariance matrix gives a final 68.3\% confidence limit of
\begin{eqnarray}
\sigma_{8} \left(
\frac{\Omega_{m}}{0.3} \right)^{0.48} \hspace{-5mm} & = & \hspace{-3mm} 0.81 \pm
0.075 \pm 0.024 \pm 0.05 \pm 0.02
\nonumber \\
  & = & \hspace{-3mm} 0.81 \pm 0.17 ~,
\label{eqn:con_full_error2d}
\end{eqnarray}
\noindent where the various systematic errors have been combined linearly on the
second line.

\begin{figure}[tb]
\epsscale{0.84}
\plotone{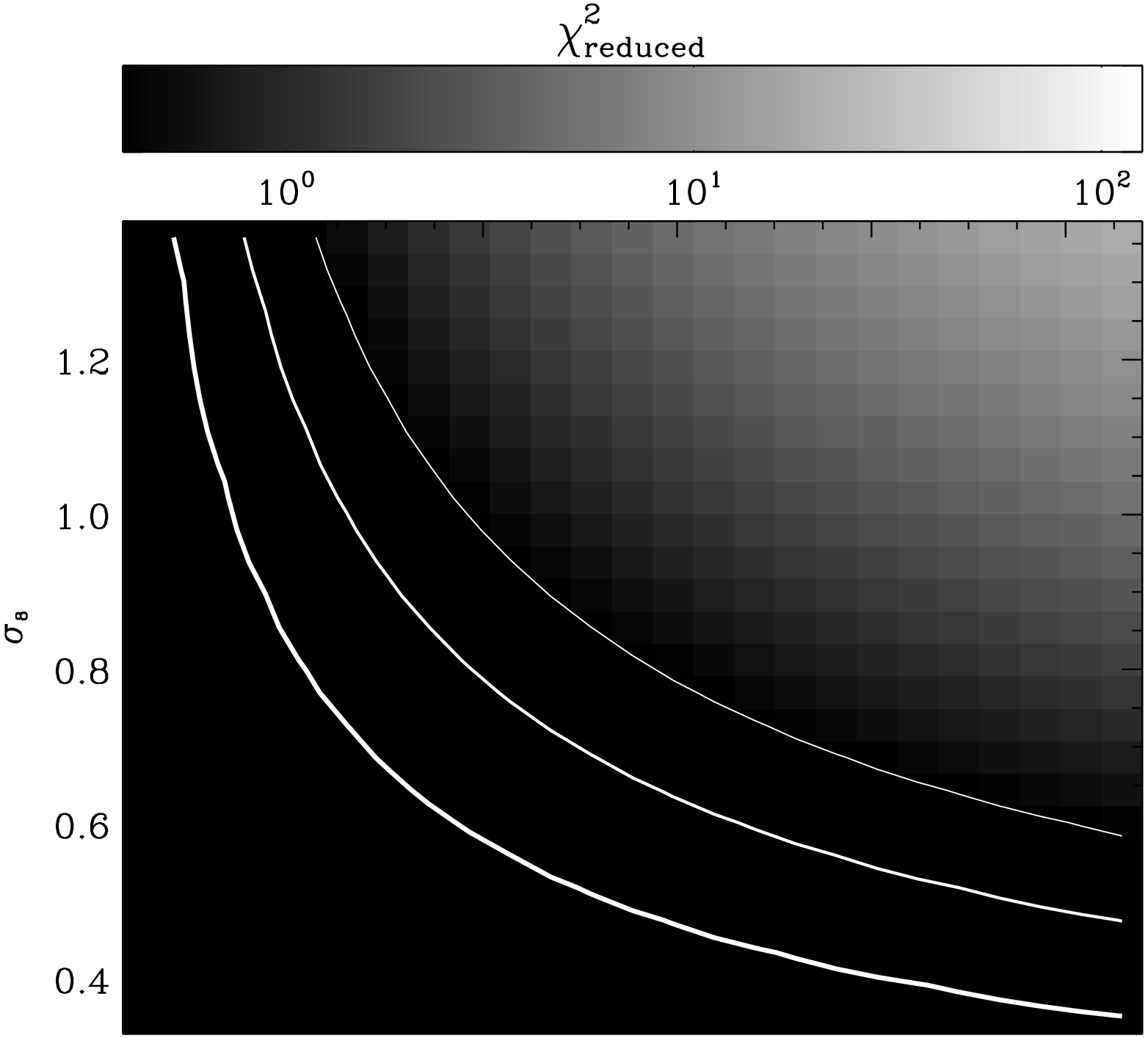}\\[5mm]
\plotone{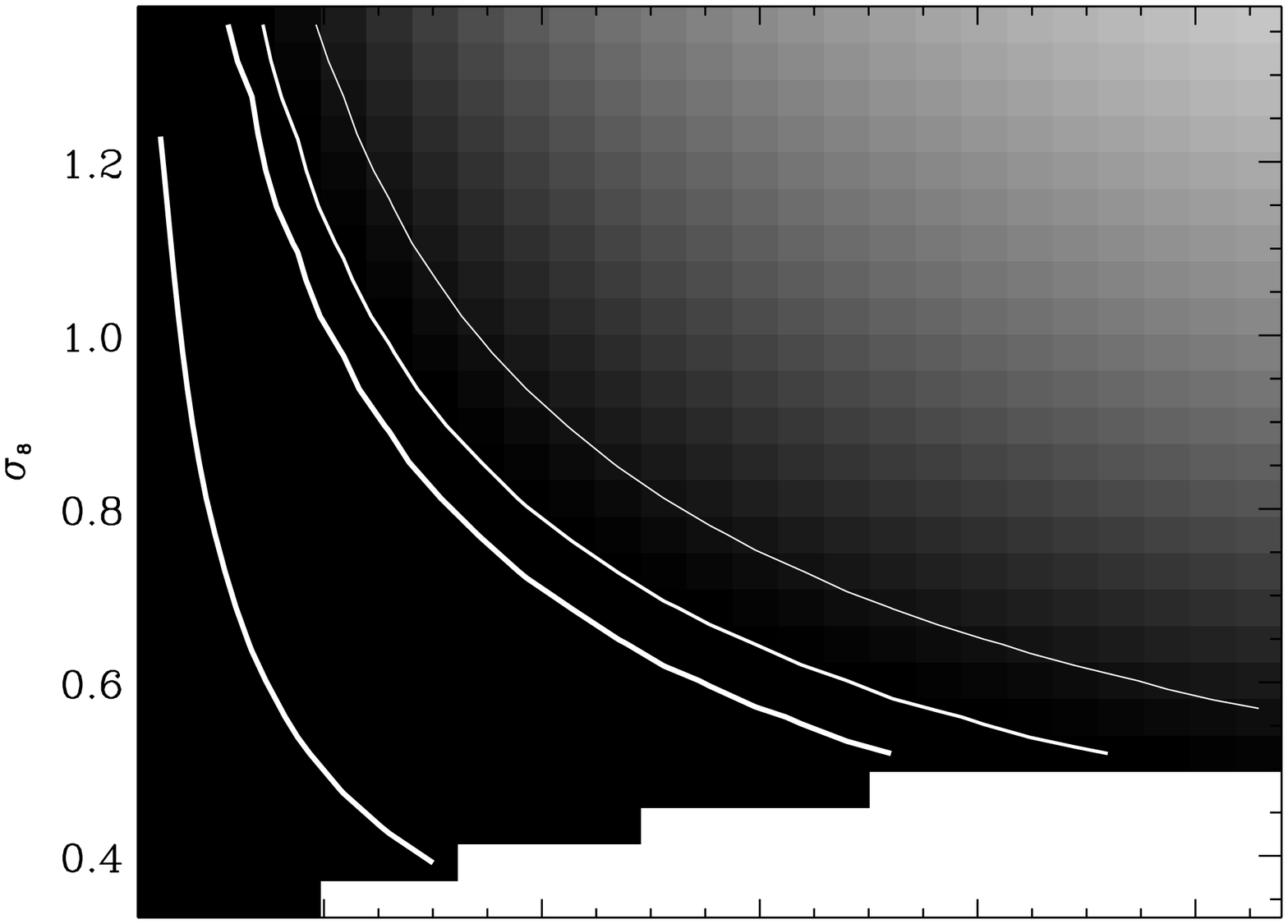}\\[5mm]
\plotone{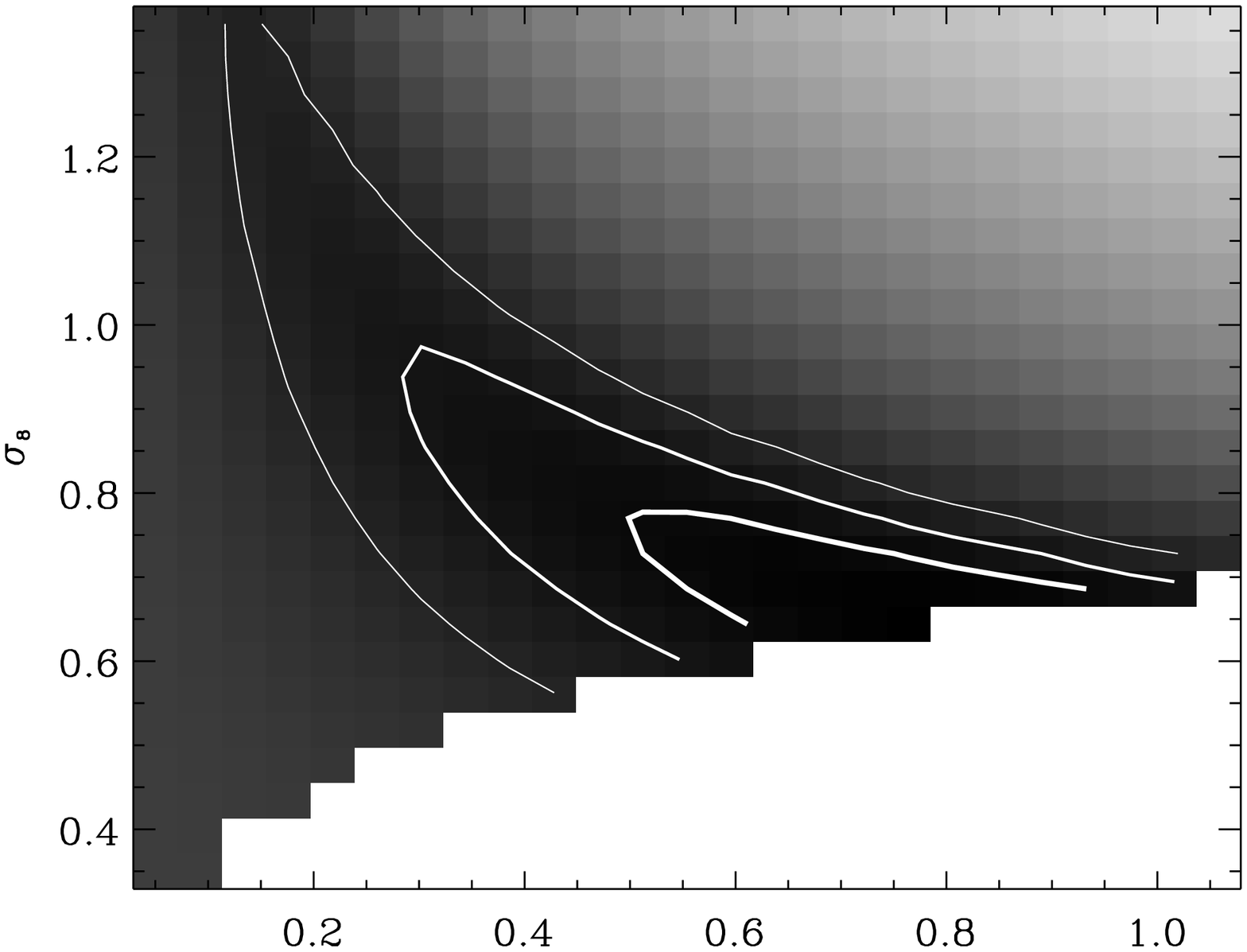}
\epsscale{1}
\caption{Constraints on cosmological parameters from within each of the three
separate redshift slices, from low (top panel) to high redshift (bottom panel). The redshift binning scheme is shown in
figure~\ref{fig:3dzdist} and discussed in the text. The contours indicate 
$68.3\%$, $95.4\%$ and $99.7\%$ confidence limits, and the logarithmic color
scale is common to all three slices.}
\label{fig:2.5ds8omslices}
\end{figure}

\subsection{3D parameter constraints} \label{sec:3dconstranits}

We shall now include the redshift information available for each object,
adopting the 3D binning scheme introduced in \S\ref{sec:3dresults}. A simple 2D
analysis can first be performed within each redshift slice, by simply exchanging
the redshift sensitivity function $g(z)$ calculated using the full redshift
distribution for one calculated using the restricted distributions.
Figure~\ref{fig:2.5ds8omslices} shows the constraints on cosmological parameters
from each slice, using only pairs of galaxies where both pairs lie in that
slice, but the full covariance matrix for each. The individual results are
clearly more noisy than for the full 2D analysis, since each slice contains only
approximately one ninth of the number of galaxy pairs. However, all of the
slices are consistent with our base cosmological model. Furthermore, while the
statistical noise is similar in each slice, because they all contain a similar
number of galaxy pairs, the signal (and hence the signal to noise) clearly
increases at high redshift, as expected.

\begin{figure}[tb]
\epsscale{0.84}
\plotone{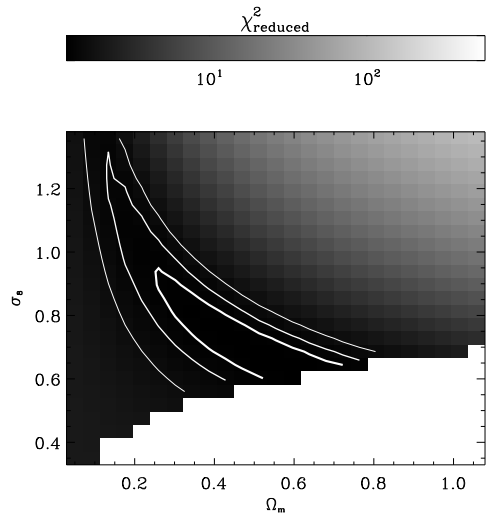}
\epsscale{1}
\caption{Combined constraints on cosmological parameters $\Omega_{\rm m}$ and
$\sigma_8$ from a series of effectively 2D shear analyses in each of the three
redshift slices (see text). Only pairs of galaxies where both lie in the same
redshift slice have been included in this analysis. This can be compared to the
similar result from the 2D analysis in figure~\ref{fig:2ds8om}, and the full 3D
analysis in figure \ref{fig:3ds8om}.}
\label{fig:2.5ds8om}
\end{figure}

In figure~\ref{fig:2.5ds8om}, the constraints from the three redshift bins are
combined as if they all provided independent information (despite the fact that
the redshift sensitivity functions in figure~\ref{fig:3dzdist} clearly overlap,
so they are correlated). Although there are approximately only
one third of the number of galaxy pairs in this analysis as there were in the 2D
analysis, the additional information about the evolution of the signal as a
function of redshift retightens the 68\% confidence limit constraints back to 
a similar value of 
\begin{equation} 
\sigma_{8} \left(\frac{\Omega_{m}}{0.3} \right)^{0.44} = 0.86 \pm 0.08 ~,
\label{eqn:2.5ds8om} \end{equation}
\noindent for $\Omega_{m}\geqslant0.25$. The best-fit model has
$\Omega_{m}=0.55$ and $\sigma_{8}=0.64$, which achieves $\chi^2_{\rm
reduced}=1.18$ in 28 degrees of freedom.

\begin{figure}[tb]
\vspace*{5mm}
\epsscale{0.84}
\plotone{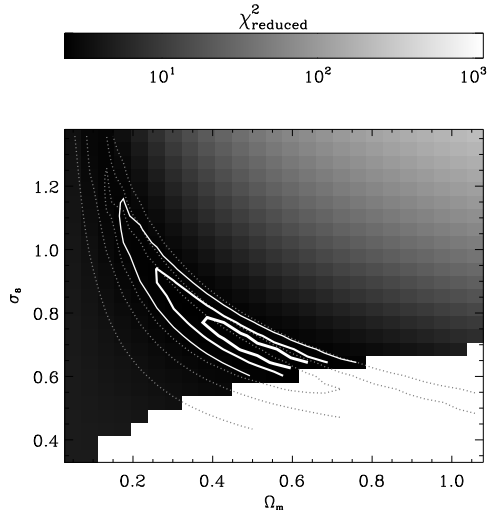}
\epsscale{1}
\caption{Constraints on cosmological parameters $\Omega_{\rm m}$ and $\sigma_8$, 
from a full 3D cosmic shear analysis. 
Solid contours indicate $68.3\%$, $95.4\%$ and $99.7\%$
confidence limits due to statistical errors and marginalization over other
parameters;
potential sources of additional, systematic error are discussed in the
text. These constraints are far tighter than the equivalent results from our
simple 2D analysis, which are reproduced from figure~\ref{fig:2ds8om} 
as dotted lines for ease of comparison. The white area at the
bottom-right was excluded because the \citet{smith} fitting functions could
not be evaluated without unreasonable extrapolation of the non-linear
matter power spectrum to physical scales smaller than $0.1h^{-1}$kpc.}
\label{fig:3ds8om}
\end{figure}

We can restore the missing galaxy pairs, and their information content, by
introducing three additional correlation functions, constructed from pairs of
galaxies that lie in different redshift slices. The theoretical expectation for
these correlation functions requires the $g^2(z)$ term in
equation~(\ref{eqn:shearpowerspectrum}) to be replaced by the product of the
lensing sensitivity functions for the two redshift bins. We use the full
covariance matrix, which is again estimated from variation between the four
quadrants of the COSMOS field. Figure~\ref{fig:3ds8om} shows a projection of the
log-likelihood surface, with the usual contours.

The best-fit model has $\Omega_{m}=0.47$ and $\sigma_{8}=0.72$, which achieves
$\chi^2_{\rm reduced}=2.35$ in 56 degrees of freedom. This is significantly greater
than unity because only statistical errors are currently included. As described
below, the error budget is increased by a factor of $\sim1.5$, and the minimum 
$\chi^2_{\rm reduced}$ to 1.04, when considering systematic errors in the relative
shear calibration and mixing of galaxies between bins. Again we find the usual
degeneracy, the best-fit position along which is determined by the parameter
$\Gamma$. However, with the full 3D information, parameter constraints in the
direction orthogonal to this are much tighter. Our 68\% confidence limits are
well-fit by
\begin{equation}
\sigma_{8} \left(\frac{\Omega_{m}}{0.3} \right)^{0.44} = 0.866 \pm 0.033 ~,
\label{eqn:3ds8om}
\end{equation}
\noindent for $0.3\leqslant\Omega_{m}\leqslant 0.6$. 

We now incorporate a systematic error budget into our 3D parameter constraints. We
allow a 6\% absolute shear calibration uncertainty \citep{apjse_lea}, a 5\%
relative shear calibration uncertainty between low and high redshift bins, and a
potential 10\% contamination \citep[\eg][]{snap2} of the high redshift bin by
galaxies really at low redshift (and vice versa) due to the possibility of
catastrophic redshift errors discussed in \S\ref{sec:photozintro}. This leaves a
final 68.3\% confidence limit of
\begin{eqnarray}
\sigma_{8} \left(
\frac{\Omega_{m}}{0.3} \right)^{0.44} \hspace{-5mm} & = & \hspace{-3mm} 
0.866 \pm 0.033 \pm 0.026 \pm 0.009 ^{+0.017}_{-0.000}
\nonumber \\
  & = & \hspace{-3mm} 0.866^{+0.085}_{-0.068} ~,
\label{eqn:con_full_error3d}
\end{eqnarray}
\noindent where the various systematic errors have been combined linearly on the
second line. Note that, when considering the relative improvement in the parameter
constraints from a 2D analysis~(\ref{eqn:con_full_error2d}) to a 3D analysis
(\ref{eqn:con_full_error3d}), it is not appropriate to include errors from
uncertainty in the absolute calibration of a shear measurement method that is common
to both. Continuing to budget for potential relative mis-calibration between low- and
high-redshift bins, as well as including all other sources of systematic and
statistical error, reveals a dramatic {\it threefold} tightening of parameter
constraints.

We have also tried increasing the number of redshift slices, for a finer quantitative
measurement of the evolution of the shear signal. We attempted an analysis using five
redshift bins, created by splitting in half the first two slices of the three used
previously. Unfortunately, the covariance matrix became degenerate, and harder to
invert. Furthermore, the best-fit$\chi^2_{\rm reduced}$ and cosmological parameter
constraints degraded. The results in each bin were very noisy (the signal to noise is
proportional to $n^{-2}_{\rm galaxies}$), but, as in \S\ref{sec:growth}, there were
hints that the signal did not evolve as expected after this finer redshift binning.
The likelihood surfaces from individual slices did not agree, so their combination
was blurred out. We interpret this as indicating that galaxies were beginning to be
placed in the wrong redshift bins, and polluting that signal. Thus we have
effectively reached the available precision of the photometric redshifts, at least at
the high redshifts in which the weak lensing signal is concentrated. For further
progress, we await ongoing, deeper multicolor observations of the COSMOS field.



\section{Conclusions} \label{sec:conclusions}

We have performed a fully three dimensional cosmic shear analysis of the largest
ever survey with the Hubble Space Telescope. The 3D shear field contains rich
information about the growth of structure and the expansion history of the
universe. Indeed, by assuming a concordance cosmological model, we have directly
measured the growth of structure on both linear and non-linear physical scales.
We have also placed independent 68\% confidence limits on cosmological
parameters. From a traditional, two dimensional cosmic shear analysis, we
measure
\begin{equation}
\sigma_{8} \left(\frac{\Omega_{m}}{0.3} \right)^{0.48} = 0.81\pm 0.17 ~,
\end{equation}
\noindent with $0.15\leqslant\Omega_{\rm m}\leqslant 0.7$.
From a full, three dimensional analysis of the same data, we obtain
\begin{equation}
\sigma_{8} \left(\frac{\Omega_{m}}{0.3} \right)^{0.44} = 0.866^{+0.085}_{-0.068} ~,
\end{equation}
\noindent with $\Omega_{m}\geqslant 0.3$. 
This represents a dramatic improvement over already remarkable constraints. In
fact, disregarding uncertainty in the absolute calibration of our shear
measurement method, which is common to both analyses, the 3D constraints
represent a {\it threefold} relative improvement in the errors from the 2D 
constraints.

A solely two-point cosmic shear analysis cannot easily isolate a measurement of just
$\Omega_m$. The degeneracy with $\sigma_8$ is broken only by the difference in signal between
large and small scales. Our best-fit value of $\Omega_m$ is slightly larger than the
measurement of $0.23\pm0.02$ from the 2dF galaxy redshift survey \citep{cole2df}.  As
discussed in \S\ref{sec:error}, undetected CTE correction residuals could potentially affect
our measurement on the very largest scale in each redshift slice; however this datum carries
very little weight because of shot noise, so this explanation is unlikely. Because high
redshift slices probe larger physical scales than low redshift slices, our measurement of
$\Omega_m$ could also be potentially biased by photometric redshift failures. After allowing
for this effect in our systematic error budget, the discrepancy in $\Omega_m$ is within
$1\sigma$ of the 2dF results, so we shall not pursue this further.

The main constraint from our data is upon $\sigma_8$. We find a value slightly larger than
that of $0.74^{+0.05}_{-0.06}$ from the 3-year WMAP data \citep{wmap3}. Our result is also
larger than most estimates of cluster abundance from $x$-ray surveys
\citep[\eg][]{bor,schuecker}, and from other recent space-based weak lensing measurements
\citep{gems_cs,gemscs2}. However, the HST GEMS survey, on which both of the latter were
based, suffers from sample variance due to its limited size, and is suspected from other
measures of containing an unusually empty portion of the universe. Furthermore, independent
measurements of $\sigma_8=0.85$ or slightly greater have recently been published by
\citet{fgas06}, from observations of the gas mass fraction in $x$-ray selected clusters;
\citet{giantarcs}, by counting the number of observed giant arcs; and \citet{Lya1} and 
\citet{LyaCMB}, with Lyman-$\alpha$ forest data. All of these measures contain
information about small-scale density fluctuations at relatively low redshift: something much
more intrinsically suited to a measurement of $\sigma_8$ than the CMB. Our results are also
remarkably consistent with those from the ground-based CFHT wide synoptic legacy survey
\citep{cfhtlsw}. Such agreement between the largest space-based and ground-based surveys
demonstrates the maturity of the field post-STEP. The combination of all these results is
therefore beginning to hint at inconsistencies in either the standard cosmological model or
in the interpretation of one or more of these methods.

With the profundity of this statement in mind, we are  careful to
realistically include all possible sources of systematic error. The dominant
contribution to the total error budget is uncertainty in the absolute
calibration of our shear measurement method. The weak lensing community is
earnestly working to improve and ascertain the reliability of various methods
through simulated images that contain a known input signal \citep{apjse_lea,
step1, step2}. 

Aside from this contribution, further exploitation of the COSMOS survey is currently
limited by two additional sources of potential systematic error. Conveniently, these
two limits currently happen to lie at a similar flux level and therefore affect a
similar population of galaxies, which we simply remove from our analysis. Since a
weak lensing measurement is concerned with the mass distribution in front of galaxies
rather than the galaxies themselves, this can be done without worries about bias.
Firstly, the in-orbit degradation of the ACS CCDs has led to inadequate charge
transfer efficiency during readout, which creates trailing of faint objects, and
mimics a weak lensing signal. In \citet{apjse_rho}, we formulated an empirical
correction scheme for the CTE effect, which works for all but the faintest galaxies;
an ongoing effort to correct CTE pixel-by-pixel in raw images should allow us to push
this limit and dramatically increase the number density of galaxies with measured
shears. Secondly, the finite number of colors available for each galaxy, and
particularly the depth in near-IR bands, limits the current accuracy of photometric
redshifts. Continuing observations with the Subaru telescope should improve their
precision. This will allow finer resolution in the redshift direction and, most
importantly, will break redshift degeneracies ubiquitous in the redshifts of faint
objects so that they can also be used.

By understanding the characteristics of effects that dominate real data,
COSMOS is proving an invaluable dry run for future, dedicated weak lensing
missions in space. We have revealed important aspects that should ideally be
minimized by hardware design and mission scheduling requirements. However, we
have also demonstrated the rich information content of the 3D shear field, and
shown a proof of concept for some of the proposed tomographic analysis
techniques that will be required to fully exploit such future data.




\acknowledgments

The HST COSMOS Treasury program was supported through NASA grant 
HST-GO-09822. The HST ACS CTE calibration program is supported through 
NASA grant HST-AR-10964. AL, AR, ES, JPK, LT and  YM were partly funded 
by the CNRS Programme National de Cosmologie. CH is supported by a CITA 
national fellowship. We thank Alan Heavens, Will High, Tom Kitching, 
John Peacock, Peter Schneider and Andy Taylor for illuminating discussions.
We thank Tony Roman, Denise Taylor, and David 
Soderblom for their assistance in planning and scheduling the extensive 
COSMOS observations. We thank the NASA IPAC/IRSA staff (Anastasia Laity, 
Anastasia Alexov, Bruce Berriman and John Good) for providing online 
archive and server capabilities for the COSMOS datasets. It is also our
pleasure to gratefully acknowledge the contributions of the entire COSMOS 
collaboration, consisting of more than 70 scientists. More information on 
the COSMOS survey is available at {\tt 
\url{http://www.astro.caltech.edu/~cosmos}}. 




{\it Facilities:} \facility{HST (ACS)}, \facility{Subaru (Suprime-Cam)}, \facility{CFHT
(Megacam)}.

\end{document}